\documentclass[sigconf]{acmart}

\makeatletter
\newcommand{\arxivcountryomitted}{\global\@ACM@countrypresenttrue}
\makeatother

\renewcommand\footnotetextcopyrightpermission[1]{}

\usepackage{array}
\usepackage{tabularx}
\usepackage{tikz}
\usepackage{graphicx}

\usepackage{placeins}
\usepackage{pgfplots}
\pgfplotsset{compat=1.18}
\usetikzlibrary{positioning,arrows.meta,shapes.geometric}
\usetikzlibrary{positioning,fit,calc}

\usetikzlibrary{backgrounds}
\definecolor{domc}{HTML}{2F5D8A}   
\definecolor{reac}{HTML}{2E7D6B}   
\definecolor{flowc}{HTML}{555555}  

\begin{document}

\title{RiskReason: A Benchmark for Verifiable Financial Risk Reasoning}

\titlenote{Accepted at the Workshop on The 3rd Workshop on LLMs and Generative AI for Finance }

\titlenote{Project repository:
\url{https://github.com/DataismLab/RiskReason-A-Benchmark}}

\author{Nazanin Yousefi}
\affiliation{\institution{Dataism Laboratory for Quantitative Finance}\arxivcountryomitted}

\author{Mohammadreza Ghafouri}
\affiliation{\institution{Independent Researcher}\arxivcountryomitted}

\author{Seyedreza Tavakoli}
\affiliation{\institution{Dataism Laboratory for Quantitative Finance}\arxivcountryomitted}

\author{MohammadJavad Gharegozlou}
\affiliation{\institution{Independent Researcher}\arxivcountryomitted}

\author{Arman Khaledian}
\affiliation{\institution{Independent Researcher}\arxivcountryomitted}

\author{Amir Amini}
\affiliation{\department{Bendheim Center for Finance}\institution{Princeton University}\arxivcountryomitted}
\affiliation{\institution{Dataism Laboratory for Quantitative Finance}\arxivcountryomitted}

\author{Eghbal Rahimikia}
\affiliation{\institution{Alliance Manchester Business School, University of Manchester}\arxivcountryomitted}
\affiliation{\institution{The Alan Turing Institute}\arxivcountryomitted}

\author{Gholamali Aminian}
\affiliation{\institution{The Alan Turing Institute}\arxivcountryomitted}

\author{Marjan Alirezaie}
\affiliation{\department{Department of Computer Science}\institution{Toronto Metropolitan University}\arxivcountryomitted}

\author{\texorpdfstring{\vspace{-5pt}Ali Habibnia}{Ali Habibnia}}
\authornote{Corresponding author: habibnia@vt.edu}
\affiliation{\department{Department of Economics}\institution{Virginia Tech}\arxivcountryomitted}
\affiliation{\institution{Dataism Laboratory for Quantitative Finance}\arxivcountryomitted}

\begin{abstract}
Robust financial-risk inference requires more than mathematical correctness. Models must accurately apply conditional rules, maintain evolving financial states, and make decisions that respect operational constraints. This work introduces \textbf{RiskReason}, a benchmark for verifiable financial-risk reasoning that spans diverse risk-management mechanisms. The benchmark includes Standard tasks, which assess individual financial mechanisms, and Long-horizon tasks, which combine multiple mechanisms through extended dependencies. Each task provides reference outcomes, formal specifications, reasoning annotations, and dependency representations to support systematic analysis of model performance. Eleven model configurations are evaluated using strict single-attempt accuracy, complemented by structural and qualitative failure analyses. The strongest configurations achieve
80.0--90.2\% overall accuracy, with several maintaining high performance on composed financial scenarios.
However, analysis of failures reveals incorrect risk
decisions despite accurate intermediate calculations,
inconsistent propagation of financial states, and
remediation actions that violate binding constraints.
In contrast, some strict evaluation failures result from
incomplete reporting rather than incorrect financial
conclusions. These findings indicate that aggregate
accuracy alone is insufficient to characterize
financial-risk reasoning reliability and motivate
evaluations that distinguish numerical correctness,
state consistency, constraint satisfaction, and
decision validity.
\end{abstract}


\makeatletter
\let\RiskReasonOriginalFnSymbol\@fnsymbol
\renewcommand{\@fnsymbol}[1]{}
\makeatother
\maketitle
\makeatletter
\let\@fnsymbol\RiskReasonOriginalFnSymbol
\makeatother
\pagestyle{plain}


\section{Introduction}
\label{sec:introduction}

Large language models (LLMs) have shown impressive results in financial question answering, numerical reasoning, and solving problems within specific domains~\cite{xie2024finben,tang2025financereasoning,zhang2025xfinbench}. Still, performing well on these tasks does not guarantee accurate reasoning when it comes to financial risk management. For example, a bank with \$50M in eligible liquid assets and \$45M in stressed net cash outflows would calculate a liquidity coverage ratio of 111.11\%. If \$8M of those assets become ineligible due to a downgrade rule, the ratio drops to 93.33\%. A model might compute the ratio correctly but misjudge the situation by not updating which assets are eligible. The core difficulty is not just in doing the math, but in properly applying financial rules as conditions change.

Financial risk management is a vital area where such reasoning matters greatly. Judging credit exposure, liquidity, collateral, leverage, and regulatory capital often depends on how financial relationships, contracts, and regulations interact~\cite{crouhy2023essentials,hull2023risk}. A change in any one amount may impact later obligations, resources, or allowed actions. As a result, strong reasoning here means more than just calculating risk measures—it demands keeping track of changing numbers, conditions, constraints, and decisions throughout the process.

While current financial benchmarks have improved how we evaluate financial knowledge, numerical reasoning, and complex problem-solving —including tasks related to risk and structured verification~\cite{xie2024finben,tang2025financereasoning,xie2026finchain}—they do not fully capture how different financial risk mechanisms work together. Specifically, knowing a model’s accuracy on overall tasks does not reveal if it can monitor changes in financial states, follow conditional rules, respect constraints, or use its calculations to make the right risk decisions.

To explore these abilities, this article presents \textbf{RiskReason}—a benchmark designed for evaluating financial risk reasoning in a verifiable way. RiskReason covers a variety of financial-risk mechanisms with two types of tasks. Standard tasks focus on individual or basic combinations of mechanisms. In contrast, long-horizon tasks bring together multiple mechanisms in longer sequences, where each step’s results affect the following financial states and rules. Every task comes with a verifiable answer, a formal outline, and organized reasoning notes that show how each step is related. These features help assess both model performance and what aspects of financial problems lead to a model’s success or failure.

This work offers three main contributions: (1) a specialized benchmark for assessing financial-risk reasoning in both individual and long-term, combined scenarios; (2) structured task outlines that clearly represent financial values, conditional rules, system changes, and their interdependencies; and (3) an evaluation of eleven model setups, looking at overall results, structural links, and common mistakes. The results show that strong model performance on complex financial scenarios can still coincide with major errors in applying rules, updating states, or meeting constraints. In addition, incomplete or unconventional reporting can lead to evaluation failures, even when the financial decision itself isn’t wrong. Altogether, these findings highlight the need for more decision-focused assessments of financial reasoning reliability, rather than just aggregate accuracy.

\section{Related Work}
\label{sec:related-work}

\subsection{Financial Reasoning Benchmarks}

Financial reasoning benchmarks have progressed from numerical question answering over financial reports and tables, as exemplified by FinQA and TAT-QA~\cite{chen2021finqa,zhu2021tatqa}, toward broader evaluations of financial question answering, domain knowledge, and complex problem solving~\cite{islam2023financebench,xie2024finben,tang2025financereasoning,zhang2025xfinbench}. Complementary initiatives such as PIXIU and Open-FinLLMs have developed finance-specialized models and evaluation resources covering diverse tasks and data modalities~\cite{xie2023pixiu,huang2024openfinllms}.

These developments have substantially expanded financial LLM evaluation. However, their primary emphasis is on broad financial capabilities, numerical reasoning, or knowledge-intensive problem solving. Although several benchmarks include risk-related tasks, the consistency of financial-risk calculations, state transitions, conditional constraints, and resulting decisions is not generally their central organizing principle.

\subsection{Financial Risk Reasoning}

Financial risk management involves more than isolated calculations. Risk outcomes can depend on contractual and regulatory conditions, threshold activation, temporal state changes, and the propagation of shocks across interconnected market, credit, liquidity, collateral, and capital positions~\cite{crouhy2023essentials,hull2023risk}. These characteristics make the consistency between computed quantities, financial states, and applicable constraints particularly important.

Research on LLMs in finance has examined stock-return
predictability from news~\cite{lopezlira2026forecast},
finance-specialized models in portfolio
management~\cite{rahimikia2024revisiting}, and
speculative behavior among LLM agents in financial
markets~\cite{cartea2026bubbles}. Ghafouri
et al.~\cite{ghafouri2025riskambiguity} further
investigate LLM risk preferences, ambiguity attitudes,
and decision-making under uncertainty. More directly
related to risk evaluation, FinBen includes
risk-management applications such as credit scoring,
fraud detection, and financial
distress~\cite{xie2024finben}, while Fazlija
et al.~\cite{fazlija2025regulatory} investigate LLM
reasoning for assigning regulatory risk weights under
the Basel III standardized approach to credit risk.

These studies establish the relevance of LLMs to financial outcomes and risk-related applications. RiskReason addresses a complementary question: whether models can consistently apply and coordinate financial-risk mechanisms to derive verifiable outcomes and decisions, particularly when intermediate changes alter subsequent obligations or constraints.

\subsection{Structured, Verifiable, and Long-Horizon Reasoning}

Recent benchmarks have introduced more explicit representations of financial reasoning. FinMathBench constructs multi-formula financial problems using hierarchical directed acyclic graphs to control computational dependencies~\cite{he2026finmathbench}. FinChain employs parameterized symbolic templates and executable reasoning traces to evaluate final-answer correctness and intermediate reasoning consistency~\cite{xie2026finchain}. V-FiLLM generates verified financial problems from executable computation trees, controlling dimensions such as computation depth, expression breadth, and financial concept complexity~\cite{larsen2026vfillm}. FinIndices further examines structural and temporal reasoning challenges across long financial statements and multi-period financial quantities~\cite{tong2026longhorizonfinance}.

RiskReason complements these approaches by focusing on dependencies among financial-risk mechanisms rather than computational composition or document length alone. Its tasks incorporate changing exposures, conditional rules, temporal transitions, and binding constraints that can determine subsequent financial states and admissible actions. Formal specifications and dependency annotations characterize these relationships, while the evaluation examines task-level performance and observable financial errors. The resulting emphasis is on the consistency of the calculation-to-decision process in financial risk management.


\section{The RiskReason Benchmark}
\label{sec:benchmark}

\begin{figure*}[t]
\centering
\includegraphics[width=0.94\textwidth,height=0.72\textheight,keepaspectratio]{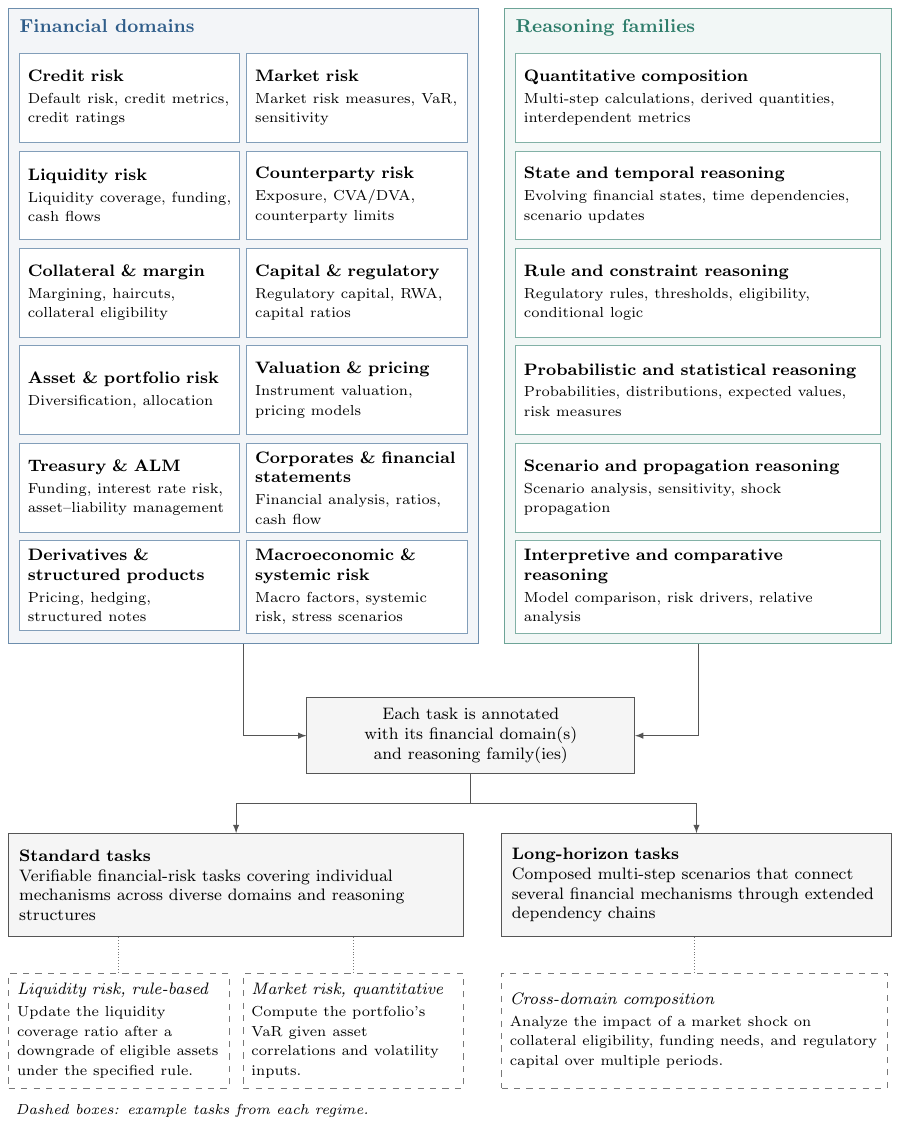}
\Description{Overview of financial risk domains and reasoning families, with examples of Standard tasks and their composition into Long-horizon tasks.}
\caption{Overview of the financial domains, reasoning families, and two task regimes of RiskReason. The examples illustrate individual mechanisms and their long-horizon composition; the annotations are not mutually exclusive.}
\label{fig:benchmark-overview}
\end{figure*}

RiskReason evaluates whether large language models
can derive verifiable financial-risk outcomes by
correctly applying financial relationships,
maintaining consistent financial states, and
respecting applicable rules and constraints.
As illustrated in Figure~\ref{fig:benchmark-overview},
the benchmark spans diverse financial domains and
reasoning requirements, extending beyond isolated
risk calculations to their interactions and
downstream consequences.

The benchmark comprises two complementary task
regimes. Standard tasks assess individual or
moderately combined financial-risk mechanisms,
whereas Long-horizon tasks connect multiple
mechanisms through extended dependencies in which
intermediate outcomes influence subsequent
financial states and decisions. Both regimes draw
on the same financial and reasoning dimensions
rather than constituting separate domain-specific
evaluations.

Each task includes a verifiable reference answer,
a formal financial specification, reasoning
annotations, and a machine-readable dependency
representation. These components support
systematic evaluation of task outcomes and
analysis of the financial relationships associated
with model successes and failures.

\subsection{Design Principles}

RiskReason follows four design principles. \textbf{Verifiability} requires outcomes that can be checked against specified financial relationships, assumptions, and constraints. When a task intentionally provides insufficient information for a unique answer, the expected outcome identifies the missing information or determinable bounds. \textbf{Mechanism orientation} emphasizes the application of financial rules, exposure and balance updates, threshold evaluation, and feasible financial decisions rather than factual recall. \textbf{Minimal-sufficient information} requires models to identify and apply the appropriate financial procedures without unnecessary solution cues. Finally, \textbf{diagnostic structure} records task-level financial and computational dependencies to support analysis beyond aggregate accuracy.

\subsection{Task Construction and Ground Truth}

Tasks are manually constructed using synthetic financial quantities, contracts, cash flows, and market or credit shocks, drawing on established financial-risk management principles~\cite{crouhy2023essentials,hull2023risk}. Rather than reproducing published exercises, scenarios are designed to combine financial relationships and constraints in controlled settings. This reduces direct overlap with existing questions, although exposure to related financial concepts during model pretraining cannot be excluded.

Reference answers are derived from the governing financial relationships, conventions, and stated constraints. Depending on the task, the required outputs may include numerical quantities, classifications, conditional decisions, multi-period financial states, or structured combinations of these elements. Intermediate quantities are documented where they determine subsequent calculations or decisions. For intentionally underdetermined tasks, the reference instead specifies the valid range or the information required to obtain a unique solution.

Task formulations and references underwent financial review for consistency and sufficiency. Identified reference and scoring issues are documented separately in the evaluation audit rather than assumed to be resolved. Where available, original and refined formulations are retained as task annotations, but are not treated as separate evaluation regimes.

\subsection{Reasoning Taxonomy and Complexity}

Each task is annotated by financial domain, subdomain, and relevant reasoning characteristics, as illustrated in Figure~\ref{fig:benchmark-overview}. These annotations are not mutually exclusive: a scenario may combine quantitative calculations, temporal dependencies, conditional rules, and shock propagation. Additional annotations describe operations, conditional logic, units, and representations.

Reasoning demands are not determined by numerical workload or prompt length alone. Tasks may involve similar calculations but differ in the dependencies among intermediate results, the activation of conditional rules, or the need to maintain financial states across successive events. RiskReason therefore combines authored reasoning annotations with graph-derived measures of task structure. These measures characterize different aspects of the problems but are not independently validated difficulty scores. Their associations with observed model performance are examined in Section~5.

\subsection{Formal Representation and Reasoning DAGs}
\label{sec:formal-dag}

Each task includes a formal specification identifying its inputs, financial relationships, conditions, constraints, and required outputs. Dependencies among these elements are encoded as a directed acyclic graph (DAG),

\begin{equation}
G_T = (V_T,E_T),
\end{equation}

where each vertex represents a task-relevant quantity or reasoning operation, and an edge $(u,v)\in E_T$ indicates that determining $v$ depends on $u$.

The machine-readable graphs distinguish five vertex types: inputs, operations, conditions, states, and outputs. Their edges capture data dependencies between quantities, control dependencies associated with conditional rules, temporal dependencies between successive states, and dependencies connecting intermediate results to requested outputs. Shared intermediate quantities can support multiple downstream operations, while state changes are represented at the relevant stages. Finite repeated calculations may be summarized through series-level vertices rather than expanded into individual iterations.

This representation captures the intended logical structure of a financial task, not a model's observed or prescribed chain of thought. It enables the characterization of structural properties such as longest-path depth, branching, merging, conditional operations, and state transitions. Because these properties depend partly on the granularity of graph construction, they describe the represented dependency structure rather than an intrinsic, representation-independent measure of task difficulty.

The formal specifications and graphs are not provided to evaluated models, which respond to the natural-language tasks. Instead, they support structural analysis and, where model responses expose relevant intermediate results, examination of observable departures from the expected financial relationships. They do not reveal unobserved model reasoning or automatically establish the cause of a failed response.

\subsection{Long-Horizon Financial Risk Reasoning}

RiskReason's Long-horizon regime extends financial-risk reasoning through the composition of dependent mechanisms rather than textual length alone (Figure~\ref{fig:benchmark-overview}). These tasks combine processes such as market-value shocks, collateral requirements, funding actions, cash-flow allocation, and capital constraints, so that the outcome of one stage changes the financial quantities or applicable conditions of subsequent stages.

Correct solutions must propagate financial states across successive events, apply conditional rules at the appropriate stages, and evaluate downstream decisions against updated balances and constraints. For example, a market shock may reduce collateral value, trigger a margin requirement, create a liquidity shortfall, and require additional borrowing that subsequently affects leverage compliance. The financial outcome depends on the consistency of the entire sequence rather than the correctness of any single calculation.

Where Long-horizon tasks incorporate mechanisms also represented in the Standard regime, their source relationships are recorded to support compositional analysis. These relationships do not constitute controlled before-and-after comparisons, as the composed scenarios may differ in parameters, conditions, and required outputs. Accordingly, comparisons between the two regimes characterize observed performance on different classes of financial problems rather than isolating the causal effect of reasoning-chain length.


\section{Evaluation Methodology}
\label{sec:evaluation}

\subsection{Models and Evaluation Scope}

We evaluate eleven model configurations: Claude Sonnet 5.5; GPT-6 Astra, Sol, and Luna; Gemini 3.1 Pro Preview and Gemini 3.8 Flash; Phi-4-Reasoning-Plus (BF16); DeepSeek-R1-Distill-Qwen-32B (Q8\_0); Fin-o1-8B (Q8\_0); DianJin-R1-7B (Q8\_0); and Qwen2.5-7B-Instruct (Q8\_0). The evaluation covers two task regimes: Standard, comprising 175 tasks, and Long-horizon, comprising 30 composed scenarios. Task identifiers retain their original prefixes for reproducibility, but the evaluation reports only the Standard and Long-horizon regimes.

Nine configurations have recorded grades for all 205 evaluated tasks. GPT-6 Astra and Sol each have 204 grades, missing \texttt{risk-119} and \texttt{risk-117}, respectively. DeepSeek and Fin-o1 have full Standard and Long-horizon coverage. For DianJin and Qwen2.5, the corrected scoring records are treated as the final versions of their respective evaluations. The exact task-ID intersection comprises 203 tasks, including all 30 Long-horizon tasks. Missing evaluations are excluded from denominators rather than treated as failures.

The results describe the recorded configurations rather than isolating architecture or training effects. Providers, serving conditions, and response instructions were not fully standardized, while the corrected grading of selected configurations uses content-based acceptance rules. These differences qualify cross-model comparisons.

\subsection{Answer Extraction and Pass@1 Scoring}

Each evaluated model--task pair has one recorded response. We report Pass@1 as strict single-attempt task accuracy:

\begin{equation}
\operatorname{Pass@1}(m)
=
\frac{1}{|\mathcal{I}_m|}
\sum_{i \in \mathcal{I}_m}
\mathbb{I}\left[g_{m,i}=\mathrm{pass}\right],
\end{equation}

where $\mathcal{I}_m$ denotes the set of tasks evaluated for model configuration $m$, and $g_{m,i}$ is the recorded task grade. A response passes only when all required answer components satisfy the reference rubric. Numerical quantities are evaluated using the grader's tolerance and precision rules, while categorical decisions, financial states, and structured outputs are assessed against their corresponding targets.

Recorded outcomes include \texttt{pass}, \texttt{partial}, \texttt{near}, \texttt{fail\_close}, and \texttt{fail}. Only \texttt{pass} contributes to strict accuracy. Missing evaluations are excluded from the denominator rather than counted as failures. Because each task was attempted once per configuration, the reported metric describes observed single-attempt performance; repeated-trial reliability and Pass@3 are not estimated.

Strict accuracy measures completion of the full answer contract, which is not necessarily equivalent to correctness of the principal financial decision. A response may reach the appropriate risk conclusion but fail because of an incorrect secondary quantity, an omitted field, or a representation mismatch. We therefore retain available field-level grades and reviewed responses for diagnostic analysis without modifying the recorded strict-pass scores.

\subsection{Structural Evaluation}

To investigate how task characteristics relate to model performance, we join task-level grades with the benchmark's dependency representations. The resulting attributes include longest-path depth, vertex and output counts, branching and merging structures, conditional operations, and state transitions. These attributes describe the represented dependencies and are not treated as independent measures of intrinsic task difficulty.

The structural analysis includes all eleven configurations, comprising 2,253 recorded responses across 205 tasks. Of these, 1,923 correspond to the 175 Standard tasks. We examine associations between structural characteristics and strict task success, including exploratory binomial logistic regressions with model-configuration and task-source fixed effects and task-clustered standard errors. The principal explanatory variables are longest-path depth and the logarithm of one plus the number of graph output vertices, allowing dependency length to be considered alongside answer-output complexity.

Because graph construction involves representational choices, including the summarization of some repeated calculations, structural measurements depend partly on graph granularity. Moreover, output count is related to the number of opportunities for failure under strict all-fields-correct grading. The analyses are therefore interpreted as associations within RiskReason rather than causal estimates of the effects of reasoning complexity.

\subsection{Qualitative Failure Analysis and Adjudication}

We complement aggregate scores with task-level examination of available model responses, intermediate calculations, final decisions, structured outputs, and grading notes. The analysis distinguishes numerical execution errors, incorrect application of financial rules or thresholds, inconsistent state transitions, infeasible remediation decisions, and discrepancies arising from answer formatting or reference specifications.

The qualitative review focuses on observable differences between model outputs and the expected financial outcomes. Where intermediate calculations are available, task dependencies help identify how an incorrect quantity or state affects subsequent decisions. These observations do not establish a model's unobserved internal reasoning process.

Review coverage varies across configurations and grading categories. Selected partial and closely graded responses received detailed adjudication, whereas not every unsuccessful response underwent equivalent review. Complete raw responses were available for Claude but not for the Gemini analyses. Consequently, failures of answer extraction cannot always be attributed to specific financial-reasoning errors, and the selected cases are not used to estimate the prevalence of failure mechanisms.

We preserve the recorded grades for quantitative comparisons while distinguishing substantive financial mistakes from reporting discrepancies and independently identified reference issues. Cases requiring additional ground-truth or scoring adjudication are documented separately in the failure analysis and appendix.

\noindent\textbf{Data and code availability.}
The RiskReason project repository for benchmark materials, model-evaluation results, and accompanying resources is available at \url{https://github.com/DataismLab/RiskReason-A-Benchmark}.

\section{Results}
\label{sec:results}

\subsection{Overall and Matched-Task Performance}

\begin{figure*}[t]
\centering
\begin{tikzpicture}
\begin{axis}[
  width=0.92\textwidth, height=7.0cm,
  xmin=0, xmax=100, ymin=-5, ymax=100,
  xtick={0,20,40,60,80,100}, ytick={0,20,40,60,80,100},
  xlabel={Standard strict task accuracy (\%)},
  ylabel={Long-horizon strict task accuracy (\%)},
  tick label style={font=\footnotesize},
  label style={font=\small},
  grid=major, grid style={black!10},
  axis line style={black!35}, every tick/.style={black!35},
  tick align=outside, clip=false,
  legend style={draw=none, fill=none, font=\footnotesize, at={(0.02,0.98)}, anchor=north west},
]
\addplot[domain=0:100, samples=2, draw=black!35, dash pattern=on 2pt off 2pt] {x};
\addlegendentry{$y=x$}
\addlegendimage{only marks,mark=*,color=domc,mark size=3pt}
\addlegendentry{Frontier/API}
\addlegendimage{only marks,mark=square*,color=reac,mark size=3pt}
\addlegendentry{Open-weight mid-scale}
\addlegendimage{only marks,mark=triangle*,color=orange!85!black,mark size=3pt}
\addlegendentry{Open-weight 7--8B}

\addplot[only marks,mark=*,mark size=4.65pt,draw=white,line width=.45pt,fill=domc,fill opacity=.86,forget plot] coordinates {(89.7,93.3)};
\addplot[only marks,mark=*,mark size=4.6pt,draw=white,line width=.45pt,fill=domc,fill opacity=.86,forget plot] coordinates {(87.9,90.0)};
\addplot[only marks,mark=*,mark size=4.58pt,draw=white,line width=.45pt,fill=domc,fill opacity=.86,forget plot] coordinates {(87.4,86.7)};
\addplot[only marks,mark=*,mark size=4.47pt,draw=white,line width=.45pt,fill=domc,fill opacity=.86,forget plot] coordinates {(82.3,90.0)};
\addplot[only marks,mark=*,mark size=4.46pt,draw=white,line width=.45pt,fill=domc,fill opacity=.86,forget plot] coordinates {(82.3,86.7)};
\addplot[only marks,mark=*,mark size=4.38pt,draw=white,line width=.45pt,fill=domc,fill opacity=.86,forget plot] coordinates {(81.7,70.0)};
\addplot[only marks,mark=square*,mark size=3.28pt,draw=white,line width=.45pt,fill=reac,fill opacity=.86,forget plot] coordinates {(46.9,33.3)};
\addplot[only marks,mark=square*,mark size=3.28pt,draw=white,line width=.45pt,fill=reac,fill opacity=.86,forget plot] coordinates {(49.7,16.7)};
\addplot[only marks,mark=square*,mark size=2.63pt,draw=white,line width=.45pt,fill=reac,fill opacity=.86,forget plot] coordinates {(33.1,3.3)};
\addplot[
  only marks,
  mark=triangle*,
  mark size=4pt,
  mark options={
    fill=orange!85!black,
    draw=black!75,
    line width=0.55pt
  },
  forget plot
] coordinates {(8.0,0.0)};

\addplot[
  only marks,
  mark=triangle*,
  mark size=4pt,
  mark options={
    fill=orange!85!black,
    draw=black!75,
    line width=0.55pt
  },
  forget plot
] coordinates {(7.4,0.0)};
\node[font=\scriptsize,anchor=west,text=black!80] at (axis cs:94.1,96.1) {Claude};
\node[font=\scriptsize,anchor=west,text=black!80] at (axis cs:89.9,92.3) {Astra};
\node[font=\scriptsize,anchor=west,text=black!80] at (axis cs:89.5,84.8) {Sol};
\node[font=\scriptsize,anchor=east,text=black!80] at (axis cs:81.2,93.0) {Luna};
\node[font=\scriptsize,anchor=east,text=black!80] at (axis cs:81.0,86.2) {Flash};
\node[font=\scriptsize,anchor=west,text=black!80] at (axis cs:83.2,69.8) {Pro};
\node[font=\scriptsize,anchor=west,text=black!80] at (axis cs:48.1,31.2) {Phi-4};
\node[font=\scriptsize,anchor=west,text=black!80] at (axis cs:51.0,16.2) {DeepSeek};
\node[font=\scriptsize,anchor=west,text=black!80] at (axis cs:34.4,5.5) {Fin-o1};
\node[
  font=\scriptsize,
  anchor=south west,
  text=black!80
] at (axis cs:9.5,3.0) {DianJin};

\node[
  font=\scriptsize,
  anchor=north east,
  text=black!80
] at (axis cs:6.0,-1.5) {Qwen};

\node[font=\scriptsize, text=black!70, anchor=west] at (rel axis cs:0.73,0.30) {Bubble area $\propto$ overall accuracy};
\draw[fill=black!14, draw=black!35] (rel axis cs:0.75,0.23) circle[radius=4.65pt];
\node[font=\scriptsize, anchor=west] at (rel axis cs:0.775,0.23) {90\%};
\draw[fill=black!14, draw=black!35] (rel axis cs:0.75,0.16) circle[radius=3.80pt];
\node[font=\scriptsize, anchor=west] at (rel axis cs:0.775,0.16) {60\%};
\draw[fill=black!14, draw=black!35] (rel axis cs:0.75,0.09) circle[radius=2.68pt];
\node[font=\scriptsize, anchor=west] at (rel axis cs:0.775,0.09) {30\%};
\end{axis}
\end{tikzpicture}
\Description{Scatter plot of Standard versus Long-horizon strict accuracy for eleven evaluated configurations. Symbol shape and color differentiate model groups; bubble area represents overall strict accuracy.}
\caption{Performance landscape across the two benchmark regimes.
Marker size reflects overall strict task accuracy, with a minimum
size for visibility. The diagonal indicates equal performance
across regimes. Points below it perform worse on Long-horizon
tasks. Qwen2.5 and DianJin achieve zero Long-horizon accuracy
and appear near the horizontal axis.}
\label{fig:performance-landscape}
\end{figure*}
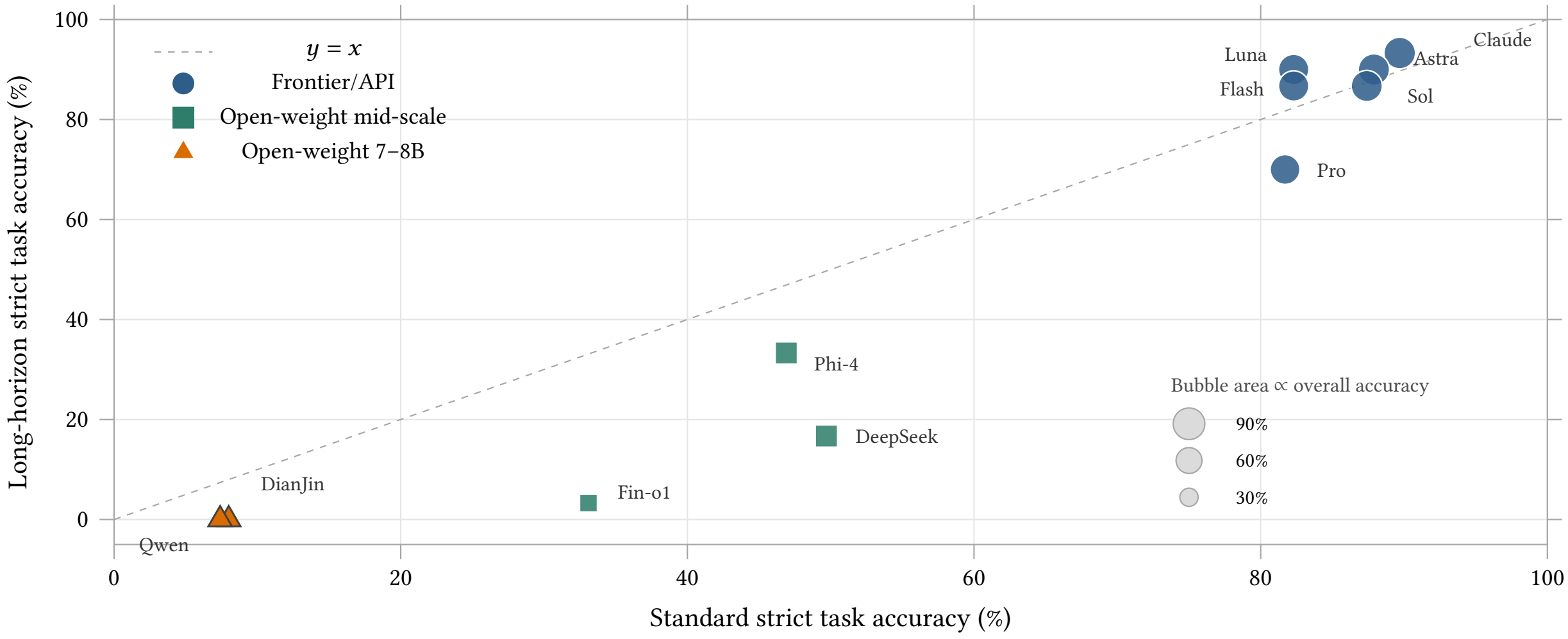

Table~\ref{tab:main-results} shows a wide range of strict single-attempt performance across the evaluated configurations. Six achieve at least 80\% overall accuracy, led by Claude Sonnet 5.5 (185/205; 90.2\%), while smaller configurations have substantially lower rates. Yet high aggregate accuracy does not amount to uniform financial reliability: even the leading model leaves one in ten tasks without a strict pass, and configurations with similar overall results can behave differently on composed scenarios. The contrast between Phi-4 and DeepSeek, each with 92/205 passes but different Long-horizon outcomes, illustrates why the two regimes and individual task outcomes must be examined alongside the aggregate score. Figure~\ref{fig:performance-landscape} visualizes this relationship by placing each configuration simultaneously in the Standard and Long-horizon regimes.

The 203 identical tasks attempted by all eleven configurations produce similar ordering at the top, with Claude passing 184, Astra 180, and Sol 178. Differences are modest: on the 204 shared Claude--Astra tasks, Claude alone passes ten and Astra alone five; against Sol these counts are ten and four. These comparisons use identical tasks and do not treat missing grades as unsuccessful attempts.

\begin{table*}[!t]
\centering
\caption{Strict single-attempt task accuracy, reported as passed/evaluated (\%). The common subset comprises 203 identical tasks attempted by all eleven configurations. Two GPT-6 runs omit one task each; missing evaluations are excluded rather than counted as failures.}
\label{tab:main-results}
\small
\setlength{\tabcolsep}{6pt}
\begin{tabular*}{\textwidth}{@{\extracolsep{\fill}}lcccc}
\toprule
\textbf{Model configuration} & \textbf{Overall} & \textbf{Standard} & \textbf{Long-horizon} & \textbf{Common (203)} \\
\midrule
\textbf{Claude Sonnet 5.5} & \textbf{185/205 (90.2\%)} & \textbf{157/175 (89.7\%)} & \textbf{28/30 (93.3\%)} & \textbf{184/203 (90.6\%)} \\
GPT-6 Astra & 180/204 (88.2\%) & 153/174 (87.9\%) & 27/30 (90.0\%) & 180/203 (88.7\%) \\
GPT-6 Sol & 178/204 (87.3\%) & 152/174 (87.4\%) & 26/30 (86.7\%) & 178/203 (87.7\%) \\
GPT-6 Luna & 171/205 (83.4\%) & 144/175 (82.3\%) & 27/30 (90.0\%) & 171/203 (84.2\%) \\
Gemini 3.8 Flash & 170/205 (82.9\%) & 144/175 (82.3\%) & 26/30 (86.7\%) & 170/203 (83.7\%) \\
Gemini 3.1 Pro Preview & 164/205 (80.0\%) & 143/175 (81.7\%) & 21/30 (70.0\%) & 164/203 (80.8\%) \\
Phi-4-Reasoning-Plus & 92/205 (44.9\%) & 82/175 (46.9\%) & 10/30 (33.3\%) & 92/203 (45.3\%) \\
DeepSeek-R1-Distill-32B & 92/205 (44.9\%) & 87/175 (49.7\%) & 5/30 (16.7\%) & 92/203 (45.3\%) \\
Fin-o1-8B & 59/205 (28.8\%) & 58/175 (33.1\%) & 1/30 (3.3\%) & 59/203 (29.1\%) \\
DianJin-R1-7B & 14/205 (6.8\%) & 14/175 (8.0\%) & 0/30 (0.0\%) & 14/203 (6.9\%) \\
Qwen2.5-7B-Instruct & 13/205 (6.3\%) & 13/175 (7.4\%) & 0/30 (0.0\%) & 13/203 (6.4\%) \\
\bottomrule
\end{tabular*}
\end{table*}

\subsection{Long-Horizon Financial Reasoning}

Long-horizon tasks assess whether models can maintain consistent financial states, apply conditional rules, and satisfy constraints across successive decisions. Despite strong performance among leading configurations, substantial differences emerge across models (Table~\ref{tab:main-results}). Notably, Phi-4 and DeepSeek achieve identical overall accuracy but differ considerably on Long-horizon tasks, indicating that aggregate performance can conceal weaknesses in composed financial reasoning.

Reviewed failures reveal several points at which these dependencies break down (Table~\ref{tab:failure-cases}). Models may correctly compute initial exposures or identify risk triggers but fail to propagate updated balances, apply rules at the appropriate stage, or verify that a proposed corrective action satisfies all constraints. Such errors can invalidate financially consequential decisions even when most intermediate calculations are correct. Conversely, some non-passes arise from reporting discrepancies or incomplete outputs rather than demonstrably incorrect financial reasoning. These observations suggest that Long-horizon reliability depends on maintaining consistency from initial calculations through final decisions, rather than reasoning length alone.

\subsection{Agreement and Task Characteristics}

Comparable overall rates can conceal different solved tasks. For example, Phi-4 and DeepSeek each pass 92 of 205 tasks, but overlap on only 64 passes, each solving 28 tasks that the other does not. Likewise, Gemini Flash and Pro pass different subsets despite their relatively close aggregate scores. On the 203-task common set, 13 tasks receive no exact pass from any configuration, while five are passed by all eleven; these counts reflect strict answer contracts as well as problem-solving performance.

Graph-derived measures provide a complementary view of task characteristics. Pooling the eleven configurations on Standard tasks, exact-pass rates are 621/957 (64.9\%) for tasks with one to three DAG output nodes and 114/242 (47.1\%) for tasks with at least ten. In an exploratory logistic regression with configuration and task-source controls and task-clustered standard errors, the number of output nodes is negatively associated with strict-pass odds (OR $\approx 0.48$ per doubling of $1+\text{output nodes}$). The adjusted longest-path measure shows no equally clear association ($p\approx 0.060$). Graph output counts include unscored nodes, and strict scoring creates more opportunities for mismatch in tasks requesting more fields; these results describe associations, not causal effects.

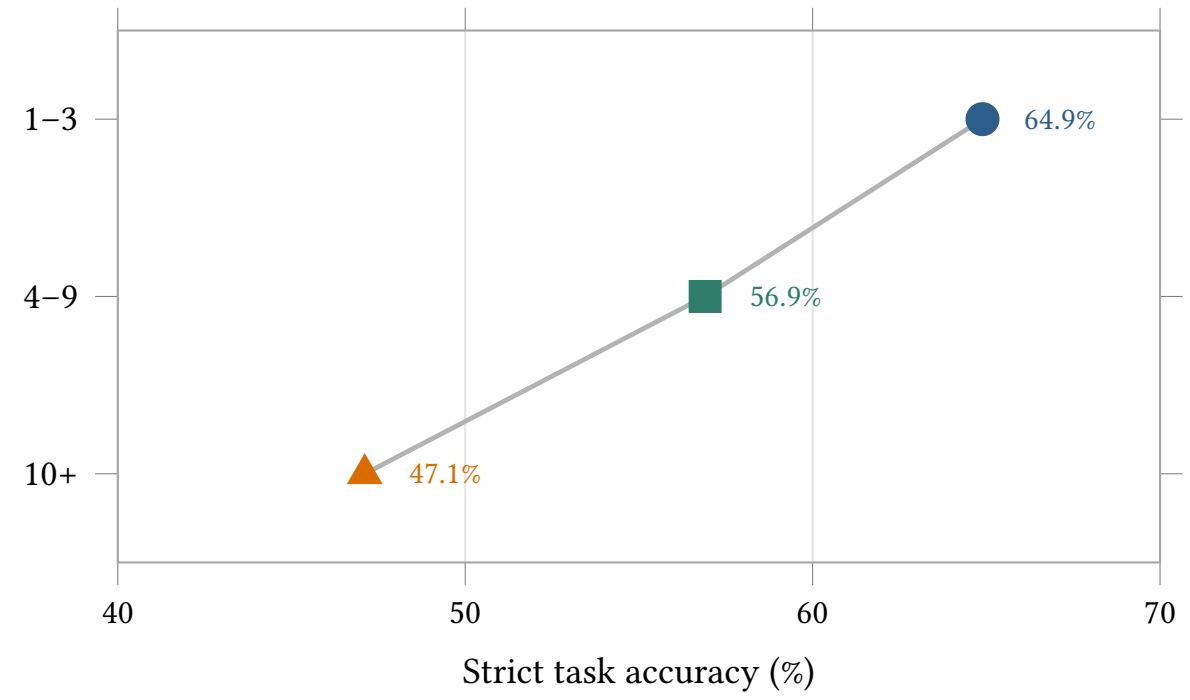
\begin{figure}[t]
\centering
\begin{tikzpicture}
\begin{axis}[
  width=\columnwidth,height=5.1cm,
  xmin=40,xmax=70,ymin=.5,ymax=3.5,y dir=reverse,
  xtick={40,50,60,70},
  ytick={1,2,3},yticklabels={1--3,4--9,10+},
  yticklabel style={font=\footnotesize},xticklabel style={font=\scriptsize},
  xlabel={Strict task accuracy (\%)},xlabel style={font=\footnotesize},
  xmajorgrids,grid style={black!10},axis line style={black!35},
  tick align=outside,clip=false,
]
\addplot[draw=black!30,line width=0.9pt] coordinates {(64.8903,1) (56.9061,2) (47.1074,3)};
\addplot[only marks,mark=*,mark size=3.0pt,color=domc] coordinates {(64.8903,1)};
\addplot[only marks,mark=square*,mark size=2.9pt,color=reac] coordinates {(56.9061,2)};
\addplot[only marks,mark=triangle*,mark size=3.5pt,color=orange!85!black] coordinates {(47.1074,3)};
\node[font=\scriptsize,anchor=west,text=domc] at (axis cs:65.6,1) {64.9\%};
\node[font=\scriptsize,anchor=west,text=reac] at (axis cs:57.7,2) {56.9\%};
\node[font=\scriptsize,anchor=west,text=orange!85!black] at (axis cs:47.9,3) {47.1\%};
\end{axis}
\end{tikzpicture}
\Description{Three-point plot showing pooled Standard-task strict accuracy decreasing from 64.9 percent on tasks with one to three graph outputs, to 56.9 percent with four to nine, to 47.1 percent with ten or more.}
\caption{Pooled Standard accuracy by DAG output count, across all eleven evaluated configurations. The three bins contain 957, 724, and 242 recorded responses, respectively. The gradient is descriptive: graph output counts are representation-dependent and create more opportunities for mismatch under strict all-fields-correct scoring.}
\label{fig:output-complexity}
\end{figure}

Figure~\ref{fig:output-complexity} summarizes the resulting gradient across all Standard evaluations. The pooled rate decreases as graph output count rises, although individual model patterns are not uniformly monotonic. Since output count is correlated with answer requirements, neither this descriptive gradient nor the adjusted regression isolates a causal effect of graph complexity.

Figure~\ref{fig:family} disaggregates Standard-task
performance across six authored financial reasoning families.
The variation in model rankings suggests that aggregate
accuracy alone does not fully characterize their relative
strengths across different financial reasoning requirements.

\begin{figure*}[t]
\centering
\begin{tikzpicture}[cell/.style={minimum width=1.42cm,minimum height=0.47cm,inner sep=0pt,font=\scriptsize,draw=white,line width=0.5pt}]
\node[anchor=south,align=center,font=\scriptsize,text width=1.40cm] at (0.00,0.57) {Quantitative\\composition\\ $n=45$};
\node[anchor=south,align=center,font=\scriptsize,text width=1.40cm] at (1.44,0.57) {Temporal and\\state\\ $n=27$};
\node[anchor=south,align=center,font=\scriptsize,text width=1.40cm] at (2.88,0.57) {Rule and\\constraint\\ $n=18$};
\node[anchor=south,align=center,font=\scriptsize,text width=1.40cm] at (4.32,0.57) {Scenario\\propagation\\ $n=11$};
\node[anchor=south,align=center,font=\scriptsize,text width=1.40cm] at (5.76,0.57) {Epistemic\\ $n=11$};
\node[anchor=south,align=center,font=\scriptsize,text width=1.40cm] at (7.20,0.57) {Probabilistic\\ $n=8$};
\node[anchor=east,font=\scriptsize] at (-0.80,0.00) {Claude Sonnet 5.5};
\node[cell,fill=domc!93,text=white] at (0.00,0.00) {93};
\node[cell,fill=domc!89,text=white] at (1.44,0.00) {89};
\node[cell,fill=domc!83,text=white] at (2.88,0.00) {83};
\node[cell,fill=domc!100,text=white] at (4.32,0.00) {100};
\node[cell,fill=domc!100,text=white] at (5.76,0.00) {100};
\node[cell,fill=domc!75,text=white] at (7.20,0.00) {75};
\node[anchor=east,font=\scriptsize] at (-0.80,-0.49) {GPT-6 Astra};
\node[cell,fill=domc!91,text=white] at (0.00,-0.49) {91};
\node[cell,fill=domc!81,text=white] at (1.44,-0.49) {81};
\node[cell,fill=domc!83,text=white] at (2.88,-0.49) {83};
\node[cell,fill=domc!100,text=white] at (4.32,-0.49) {100};
\node[cell,fill=domc!91,text=white] at (5.76,-0.49) {91};
\node[cell,fill=domc!88,text=white] at (7.20,-0.49) {88};
\node[anchor=east,font=\scriptsize] at (-0.80,-0.98) {GPT-6 Sol};
\node[cell,fill=domc!93,text=white] at (0.00,-0.98) {93};
\node[cell,fill=domc!85,text=white] at (1.44,-0.98) {85};
\node[cell,fill=domc!89,text=white] at (2.88,-0.98) {89};
\node[cell,fill=domc!91,text=white] at (4.32,-0.98) {91};
\node[cell,fill=domc!82,text=white] at (5.76,-0.98) {82};
\node[cell,fill=domc!88,text=white] at (7.20,-0.98) {88};
\node[anchor=east,font=\scriptsize] at (-0.80,-1.47) {GPT-6 Luna};
\node[cell,fill=domc!87,text=white] at (0.00,-1.47) {87};
\node[cell,fill=domc!74,text=white] at (1.44,-1.47) {74};
\node[cell,fill=domc!67,text=white] at (2.88,-1.47) {67};
\node[cell,fill=domc!91,text=white] at (4.32,-1.47) {91};
\node[cell,fill=domc!91,text=white] at (5.76,-1.47) {91};
\node[cell,fill=domc!88,text=white] at (7.20,-1.47) {88};
\node[anchor=east,font=\scriptsize] at (-0.80,-1.96) {Gemini 3.8 Flash};
\node[cell,fill=domc!82,text=white] at (0.00,-1.96) {82};
\node[cell,fill=domc!70,text=white] at (1.44,-1.96) {70};
\node[cell,fill=domc!83,text=white] at (2.88,-1.96) {83};
\node[cell,fill=domc!91,text=white] at (4.32,-1.96) {91};
\node[cell,fill=domc!91,text=white] at (5.76,-1.96) {91};
\node[cell,fill=domc!88,text=white] at (7.20,-1.96) {88};
\node[anchor=east,font=\scriptsize] at (-0.80,-2.45) {Gemini 3.1 Pro Preview};
\node[cell,fill=domc!78,text=white] at (0.00,-2.45) {78};
\node[cell,fill=domc!67,text=white] at (1.44,-2.45) {67};
\node[cell,fill=domc!83,text=white] at (2.88,-2.45) {83};
\node[cell,fill=domc!100,text=white] at (4.32,-2.45) {100};
\node[cell,fill=domc!100,text=white] at (5.76,-2.45) {100};
\node[cell,fill=domc!88,text=white] at (7.20,-2.45) {88};
\node[anchor=east,font=\scriptsize] at (-0.80,-2.94) {Phi-4-Reasoning-Plus};
\node[cell,fill=domc!40,text=black] at (0.00,-2.94) {40};
\node[cell,fill=domc!41,text=black] at (1.44,-2.94) {41};
\node[cell,fill=domc!39,text=black] at (2.88,-2.94) {39};
\node[cell,fill=domc!55,text=white] at (4.32,-2.94) {55};
\node[cell,fill=domc!36,text=black] at (5.76,-2.94) {36};
\node[cell,fill=domc!62,text=white] at (7.20,-2.94) {62};
\node[anchor=east,font=\scriptsize] at (-0.80,-3.43) {DeepSeek-R1-Distill-32B};
\node[cell,fill=domc!47,text=black] at (0.00,-3.43) {47};
\node[cell,fill=domc!41,text=black] at (1.44,-3.43) {41};
\node[cell,fill=domc!56,text=white] at (2.88,-3.43) {56};
\node[cell,fill=domc!55,text=white] at (4.32,-3.43) {55};
\node[cell,fill=domc!45,text=black] at (5.76,-3.43) {45};
\node[cell,fill=domc!50,text=white] at (7.20,-3.43) {50};
\node[anchor=east,font=\scriptsize] at (-0.80,-3.92) {Fin-o1-8B};
\node[cell,fill=domc!31,text=black] at (0.00,-3.92) {31};
\node[cell,fill=domc!30,text=black] at (1.44,-3.92) {30};
\node[cell,fill=domc!28,text=black] at (2.88,-3.92) {28};
\node[cell,fill=domc!36,text=black] at (4.32,-3.92) {36};
\node[cell,fill=domc!18,text=black] at (5.76,-3.92) {18};
\node[cell,fill=domc!50,text=white] at (7.20,-3.92) {50};
\node[anchor=east,font=\scriptsize] at (-0.80,-4.41) {DianJin-R1-7B};
\node[cell,fill=domc!9,text=black] at (0.00,-4.41) {9};
\node[cell,fill=domc!6,text=black] at (1.44,-4.41) {0};
\node[cell,fill=domc!6,text=black] at (2.88,-4.41) {0};
\node[cell,fill=domc!9,text=black] at (4.32,-4.41) {9};
\node[cell,fill=domc!6,text=black] at (5.76,-4.41) {0};
\node[cell,fill=domc!12,text=black] at (7.20,-4.41) {12};
\node[anchor=east,font=\scriptsize] at (-0.80,-4.90) {Qwen2.5-7B-Instruct};
\node[cell,fill=domc!6,text=black] at (0.00,-4.90) {4};
\node[cell,fill=domc!6,text=black] at (1.44,-4.90) {4};
\node[cell,fill=domc!6,text=black] at (2.88,-4.90) {0};
\node[cell,fill=domc!9,text=black] at (4.32,-4.90) {9};
\node[cell,fill=domc!6,text=black] at (5.76,-4.90) {0};
\node[cell,fill=domc!12,text=black] at (7.20,-4.90) {12};
\end{tikzpicture}
\Description{Heatmap of exact pass percentages by authored primary reasoning family on 120 Standard tasks with comparable family annotations, for eleven model configurations.}
\caption{Strict pass@1 (\%) for 120 Standard tasks with authored primary-family labels, grouped by reasoning family. This is an annotated subset of the 175-task Standard regime, not a separate evaluation set. Family sizes are shown above the columns; small groups should be interpreted descriptively.}
\label{fig:family}
\end{figure*}

\section{Failure Analysis}
\label{sec:failure-analysis}

Aggregate accuracy does not distinguish incorrect financial calculations from failures in rule application, state propagation, or decision-making. Nor does every non-passing response necessarily indicate incorrect financial reasoning, as strict scoring also penalizes incomplete or noncanonical outputs. We therefore examine reviewed model responses according to the financial consequences of their observable discrepancies. Table~\ref{tab:failure-cases} summarizes representative cases, which illustrate distinct failure mechanisms and evaluation boundaries rather than their prevalence across models.

\begin{table*}[t]
\centering
\caption{Representative financial-risk reasoning errors and evaluation boundaries from reviewed model outputs. Cases illustrate distinct mechanisms and do not estimate their frequency across the benchmark.}
\label{tab:failure-cases}
\small
\renewcommand{\arraystretch}{1.15}
\begin{tabularx}{\textwidth}{@{}p{0.20\textwidth}p{0.22\textwidth}X@{}}
\toprule
\textbf{Mechanism} &
\textbf{Model / Task} &
\textbf{Observed discrepancy} \\
\midrule

Decision after correct calculation &
Gemini Pro, \texttt{probe-036} &
Computes stressed PD (23.62\%) but assesses the 15\% monitoring threshold using posterior PD (13.39\%), incorrectly reporting no breach. \\

Rule activation &
Phi-4, \texttt{risk-057} &
Applies a covenant-triggered interest-spread adjustment in the wrong quarter. \\

State propagation &
Phi-4, \texttt{long-012} &
Increases rather than amortizes senior debt following a \$12M principal collection, distorting subsequent interest allocation. \\

Infeasible remediation &
Phi-4, \texttt{long-029} &
Proposed asset sales leave CET1/RWA near 5.61\%, below the required 6\% threshold. \\

Material terminal error &
DianJin, \texttt{long-018} &
Matches 24 of 25 graded fields but reports a \$2.4M permissible distribution instead of \$0.4M. \\

Branch-specific computation &
Gemini Flash, \texttt{risk-018} &
Correctly computes the non-reinvestment return (549.39\%) but reports 625.14\% rather than 603.56\% for the reinvestment scenario. \\

Optimization branch &
Claude, \texttt{risk-119} &
Obtains the long-only minimum-variance portfolio but reports incorrect unconstrained portfolio weights. \\

Answer-contract mismatch &
Claude and both Gemini models, \texttt{long-025} &
The central numerical borrowing-base and liquidity states are correct, but certain event descriptions fail the structured reporting contract. \\

Estimation execution &
Claude and both Gemini models, \texttt{risk-120} &
Claude and Gemini Pro return no computed AIC/BIC estimates; Gemini Flash reports numerical criteria that disagree with the stored reference, which remains subject to independent verification. \\

\bottomrule
\end{tabularx}
\end{table*}
\subsection{Financial Measurement and Decision Consistency}

Correctly computing a risk measure does not guarantee that a model applies it to the appropriate financial decision. In \texttt{probe-036}, Gemini 3.1 Pro Preview derives the posterior and stressed default probabilities, loss-given-default rate, and expected credit loss, but evaluates the monitoring threshold using the posterior rather than the stressed probability. Similarly, in \texttt{probe-004}, GPT-6 Sol and Luna calculate declining CET1 capital while accepting a claim of organic capital strengthening. In both cases, the relevant financial information is available, yet the resulting assessment is inconsistent with it.

The converse also occurs. In \texttt{probe-030}, Luna reaches the required escalation decision despite understating exposure at default following an incorrect collateral conversion. These cases distinguish measurement accuracy from decision accuracy: a correct risk classification can coexist with an incorrect exposure estimate, just as an accurate calculation can lead to an incorrect financial conclusion.

\subsection{Temporal Consistency and Feasible Remediation}

Sequential financial scenarios require models to update states and apply rules at the appropriate stages. In \texttt{risk-057}, reviewed Phi-4 and 7B responses mis-time a covenant-triggered interest-spread adjustment. In \texttt{long-012}, Phi-4 correctly identifies the initial cash-flow waterfall and trigger but treats a principal collection as an increase in senior debt rather than a reduction. This incorrect balance subsequently affects interest allocation, illustrating how a state-transition error can propagate through later calculations.

Other failures arise when models identify a risk constraint but fail to produce a compliant action. In \texttt{long-029}, Phi-4 recommends an asset-sale combination that leaves the CET1 ratio below the required 6\% minimum. In \texttt{long-018}, DianJin matches 24 of 25 graded fields but overstates the final permissible distribution by \$2M. These examples demonstrate that the financial significance of an error depends not only on the number of correct intermediate quantities, but also on whether the resulting action satisfies binding constraints.

\subsection{Numerical Execution and Branch Consistency}

Some responses identify the appropriate procedure but fail to execute all of its components correctly. In \texttt{risk-018}, Gemini 3.8 Flash calculates the return without dividend reinvestment correctly but produces an incorrect result for the reinvestment scenario. In \texttt{risk-116}, Gemini 3.1 Pro Preview derives the daily price-range components but substitutes a single-session range for the requested multi-period average true range. These cases illustrate that successful execution of one calculation does not guarantee consistency across related computational branches.

Claude provides an additional contrast in \texttt{risk-119}: it obtains the long-only portfolio solution but not the unconstrained minimum-variance weights. In \texttt{risk-120}, it leaves all 16 requested VAR information criteria uncomputed. In \texttt{risk-116}, its ATR of 50.2714 matches a simple moving average of true ranges, whereas the reference is approximately 50.81 under Wilder smoothing; the task requires clarification of the intended method before assigning that discrepancy wholly to model error.

Such errors differ from failures in financial interpretation: the model may select the relevant formula or intermediate quantities but miscalculate, aggregate, or report a required result. Distinguishing these mechanisms is important because an incorrect numerical result and an incorrect decision based on an otherwise correct calculation represent different limitations.

\subsection{Answer Fidelity and Reference Adjudication}

Strict task accuracy also captures discrepancies that do not necessarily establish incorrect financial reasoning. In \texttt{long-025}, Claude and both Gemini configurations preserve the principal numerical borrowing-base, cash, and revolver states in the reviewed outputs, while some event descriptions are missing or differ from the reference. Claude's \texttt{long-028} response likewise preserves the financial waterfall and capital effects, with non-passes arising from signs and percentage representations. Other responses contain alternative signs, scales, labels, or omitted structured fields. These discrepancies violate the recorded answer contract but should be distinguished from incorrect financial decisions. Similarly, an unparsed response establishes unsuccessful extraction under the evaluation procedure, not a specific reasoning failure without access to the raw response.

Reference validity introduces a separate consideration. For example, \texttt{risk-005} permits alternative median conventions, while \texttt{risk-092} contains conflicting leverage definitions. The reference estimates for \texttt{risk-120}, a vector-autoregression lag-selection task, also require independent replication. Although none of the eleven configurations receives a strict pass on that task, disagreement with an unresolved reference cannot by itself establish a numerical reasoning error. Additional numerical checks have flagged \texttt{risk-117} and \texttt{risk-125}. Claude's \texttt{risk-126} response was truncated at its 16,384-token output limit, so that non-pass is a completion-budget event rather than a localizable financial calculation error.


Taken together, the reviewed cases distinguish three aspects of financial-risk reliability: computing the required quantities, maintaining the financial states and constraints on which those quantities depend, and deriving valid financial decisions. Errors can arise at any of these stages, while some strict non-passes reflect reporting or reference issues instead. Because review coverage differs across configurations and grading categories, the selected examples identify observable failure mechanisms without establishing their relative frequencies.

\section{Discussion}
\label{sec:discussion}

Our findings highlight a distinction between numerical competence and reliable financial-risk reasoning. Several evaluated configurations achieve high accuracy, including on composed scenarios involving interacting financial mechanisms. Nevertheless, the reviewed failures show that models can compute relevant risk quantities while applying rules to incorrect states, propagating inconsistent balances, or proposing actions that violate binding constraints. These errors suggest that reliable financial-risk reasoning depends on the consistency of the entire calculation-to-decision process, rather than the correctness of individual computations alone.

Claude provides an informative high-performing contrast: 28/30 Long-horizon passes coexist with a failure to complete VAR information-criterion estimation on \texttt{risk-120} and an incorrect unconstrained optimization branch on \texttt{risk-119}. Thus, strong accuracy on extended financial compositions does not guarantee uniform competence across statistical estimation and financial-method selection.

This distinction also has implications for benchmark evaluation. Strict task accuracy is useful for assessing whether models satisfy complete answer requirements, but it combines errors with different financial consequences. A response may preserve the correct risk assessment while omitting a required field, whereas another may reproduce nearly every intermediate quantity yet reach an incorrect compliance decision or permissible action. Evaluating financial-risk reasoning therefore benefits from distinguishing numerical accuracy, state consistency, constraint satisfaction, decision validity, and reporting completeness. These dimensions are complementary and should not be reduced to an unweighted count of correct fields.

The results do not support a simple relationship between reasoning length and observed performance. Several higher-performing configurations maintain comparable accuracy across Standard and Long-horizon tasks, while others experience substantial declines. Similarly, differences in output requirements complicate the interpretation of task-level accuracy under strict scoring. These observations suggest that financial reasoning difficulty cannot be adequately characterized by task length or dependency structure alone. The nature of the financial relationships being coordinated, the conditions governing state transitions, and the validity of downstream decisions are also important considerations. Because the task regimes are not controlled matched pairs, these comparisons do not establish the causal effect of reasoning-chain length.

Cross-model comparisons further illustrate the limits of aggregate rankings. Configurations with similar accuracy can succeed on different tasks, while differences in performance cannot be attributed solely to model architecture, size, or financial specialization. Inference conditions and response instructions were not fully standardized, and the evaluation contains only one recorded attempt per model--task pair. Controlled comparisons and repeated evaluations would be necessary to separate these factors from underlying reasoning capabilities.

Taken together, RiskReason supports a more decision-oriented approach to financial LLM evaluation. Beyond checking final numerical answers, evaluations should examine whether intermediate quantities, evolving financial states, conditional rules, and resulting actions remain mutually consistent. The present findings are limited to synthetic, self-contained scenarios and do not establish reliability in operational risk-management settings. They nevertheless identify concrete failure mechanisms and evaluation dimensions that can inform future financial reasoning benchmarks and verification procedures.

\section{Limitations}
\label{sec:limitations}

\noindent\textbf{Evaluation conditions and reproducibility.}
Our findings are based on one recorded response per model--task pair and therefore do not establish repeated-trial reliability. Inference providers, serving conditions, and response instructions were not fully standardized, while some configurations were scored with corrected content-based review. Consequently, observed performance differences cannot be attributed solely to model architecture, scale, or financial specialization. Controlled evaluations with matched inference conditions and repeated attempts would provide stronger evidence about these factors.

\noindent\textbf{Scoring and reference validity.}
Strict task accuracy requires every specified output to satisfy the reference rubric, combining substantive financial errors with omissions and representation differences. The qualitative review distinguishes these outcomes where evidence permits, but coverage varies across models and grading categories, and complete raw Gemini responses were unavailable. Selected reference values and financial conventions also require independent adjudication. Claude's \texttt{risk-126} was cut off at its 16,384-token response limit, and its run log requests follow-up even though the grade summary lists no reruns; this result is retained as recorded pending an appropriately budgeted rerun. We retain the recorded grades and document these cases rather than applying unverified corrections. Additionally, the dependency graphs reflect authored task structure rather than models' internal reasoning, and their structural measurements depend partly on representation granularity.

\noindent\textbf{Benchmark scope and generalizability.}
RiskReason uses synthetic, self-contained, text-based financial scenarios. It does not capture the full complexity of operational risk management, including incomplete external data, evolving regulations, institutional controls, and tool-assisted workflows. The evaluation also does not include a matched human-expert performance baseline. Although the reviewed cases identify consequential reasoning failures, their frequencies cannot be inferred from the nonuniformly reviewed sample. Accordingly, benchmark performance should not be interpreted as evidence of suitability for autonomous financial-risk decisions.

\section{Conclusion}
\label{sec:conclusion}

We introduced RiskReason, a benchmark for evaluating verifiable financial-risk reasoning through the application of financial relationships, evolving states, and conditional constraints. Our evaluation of eleven model configurations shows that strong task accuracy, including on composed financial scenarios, can coexist with consequential errors in rule application, state propagation, and remediation decisions. Conversely, some strict evaluation failures reflect incomplete or noncanonical reporting rather than incorrect financial conclusions. These findings highlight the limitations of aggregate accuracy for characterizing financial-risk reliability and motivate evaluations that distinguish numerical correctness, state consistency, constraint satisfaction, and decision validity. By combining verifiable financial scenarios with structured task representations and qualitative failure analysis, RiskReason provides a framework for examining where financial reasoning succeeds and where it becomes unreliable. Future work should strengthen reference validation, standardize evaluation conditions, and extend assessment to repeated trials and realistic financial-risk workflows.

\section*{Declarations}

\paragraph*{Funding and computational resources.} This study received no project-specific research funding. No computing resources from Virginia Tech or the authors' other affiliated universities were used in this study.

\bibliographystyle{ACM-Reference-Format}
\bibliography{references}


\begin{thebibliography}{19}


\ifx \showCODEN    \undefined \def \showCODEN     #1{\unskip}     \fi
\ifx \showISBNx    \undefined \def \showISBNx     #1{\unskip}     \fi
\ifx \showISBNxiii \undefined \def \showISBNxiii  #1{\unskip}     \fi
\ifx \showISSN     \undefined \def \showISSN      #1{\unskip}     \fi
\ifx \showLCCN     \undefined \def \showLCCN      #1{\unskip}     \fi
\ifx \shownote     \undefined \def \shownote      #1{#1}          \fi
\ifx \showarticletitle \undefined \def \showarticletitle #1{#1}   \fi
\ifx \showURL      \undefined \def \showURL       {\relax}        \fi
\providecommand\bibfield[2]{#2}
\providecommand\bibinfo[2]{#2}
\providecommand\natexlab[1]{#1}
\providecommand\showeprint[2][]{arXiv:#2}
\makeatletter
\@ifundefined{NAT@parse@date}{}{\let\NAT@parse@date@orig\NAT@parse@date}
\@ifundefined{NAT@parse@date}{}{\def\NAT@parse@date#1#2#3#4#5#6@@{\NAT@parse@date@orig#1#2#3#4#5#6@@\def\NAT@tempyear{0000}\def\NAT@tempexlab{{?}}\ifx\NAT@year\NAT@tempyear\ifx\NAT@exlab\NAT@tempexlab\def\NAT@date{[n.\,d.]}\else\edef\NAT@date{[n.\,d.]\NAT@exlab}\fi\fi}}
\makeatother

\bibitem[Cartea et~al\mbox{.}(2026)]%
        {cartea2026bubbles}
\bibfield{author}{\bibinfo{person}{{\'{A}}lvaro Cartea},
  \bibinfo{person}{Patrick Chang}, \bibinfo{person}{Nan Chen}, {and}
  \bibinfo{person}{Mingyue Zhong}.} \bibinfo{year}{2026}\natexlab{}.
\newblock \bibinfo{booktitle}{\emph{{AI} Bubbles with Large Language Models}}.
\newblock \bibinfo{type}{{T}echnical {R}eport}. \bibinfo{institution}{SSRN}.
\newblock
\href{https://doi.org/10.2139/ssrn.6426761}{doi:\nolinkurl{10.2139/ssrn.6426761}}
\newblock
\shownote{Working paper; revised July 2026}.


\bibitem[Chen et~al\mbox{.}(2021)]%
        {chen2021finqa}
\bibfield{author}{\bibinfo{person}{Zhiyu Chen}, \bibinfo{person}{Wenhu Chen},
  \bibinfo{person}{Charese Smiley}, \bibinfo{person}{Sameena Shah},
  \bibinfo{person}{Iana Borova}, \bibinfo{person}{Dylan Langdon},
  \bibinfo{person}{Reema Moussa}, \bibinfo{person}{Matt Beane},
  \bibinfo{person}{Ting-Hao Huang}, \bibinfo{person}{Bryan Routledge}, {and}
  \bibinfo{person}{William~Yang Wang}.} \bibinfo{year}{2021}\natexlab{}.
\newblock \showarticletitle{{FinQA}: A Dataset of Numerical Reasoning over
  Financial Data}. In \bibinfo{booktitle}{\emph{Proceedings of the 2021
  Conference on Empirical Methods in Natural Language Processing}}.
  \bibinfo{publisher}{Association for Computational Linguistics},
  \bibinfo{address}{Online and Punta Cana, Dominican Republic},
  \bibinfo{pages}{3697--3711}.
\newblock
\showeprint[arxiv]{2109.00122}
\href{https://doi.org/10.18653/v1/2021.emnlp-main.300}{doi:\nolinkurl{10.18653/v1/2021.emnlp-main.300}}


\bibitem[Crouhy et~al\mbox{.}(2023)]%
        {crouhy2023essentials}
\bibfield{author}{\bibinfo{person}{Michel Crouhy}, \bibinfo{person}{Dan Galai},
  {and} \bibinfo{person}{Robert Mark}.} \bibinfo{year}{2023}\natexlab{}.
\newblock \bibinfo{booktitle}{\emph{The Essentials of Risk Management}
  (\bibinfo{edition}{3} ed.)}.
\newblock \bibinfo{publisher}{McGraw Hill}, \bibinfo{address}{New York, NY,
  USA}.
\newblock
\showISBNx{978-1264258864}


\bibitem[Fazlija et~al\mbox{.}(2025)]%
        {fazlija2025regulatory}
\bibfield{author}{\bibinfo{person}{Bledar Fazlija}, \bibinfo{person}{Meriton
  Ibraimi}, \bibinfo{person}{Aynaz Forouzandeh}, {and} \bibinfo{person}{Arber
  Fazlija}.} \bibinfo{year}{2025}\natexlab{}.
\newblock \showarticletitle{Reasoning with financial regulatory texts via Large
  Language Models}.
\newblock \bibinfo{journal}{\emph{Journal of Behavioral and Experimental
  Finance}}  \bibinfo{volume}{47} (\bibinfo{date}{Sept.} \bibinfo{year}{2025}),
  \bibinfo{pages}{101067}.
\newblock
\href{https://doi.org/10.1016/j.jbef.2025.101067}{doi:\nolinkurl{10.1016/j.jbef.2025.101067}}


\bibitem[Ghafouri et~al\mbox{.}(2025)]%
        {ghafouri2025riskambiguity}
\bibfield{author}{\bibinfo{person}{Mohammadreza Ghafouri},
  \bibinfo{person}{Nazanin Yousefi}, \bibinfo{person}{Arian Akbari},
  \bibinfo{person}{Reza Tavakoli}, \bibinfo{person}{Farbod Davoodi},
  \bibinfo{person}{Gholamali Aminian}, \bibinfo{person}{Nariman Khaledian},
  {and} \bibinfo{person}{Arman Khaledian}.} \bibinfo{year}{2025}\natexlab{}.
\newblock \bibinfo{title}{Risk, Ambiguity, and Infinity: Behavioral Signatures
  of Modern Large Language Models}.
\newblock
\href{https://doi.org/10.2139/ssrn.5688064}{doi:\nolinkurl{10.2139/ssrn.5688064}}
\newblock
\shownote{AI for Finance Symposium, ICAIF 2025, Singapore}.


\bibitem[He et~al\mbox{.}(2026)]%
        {he2026finmathbench}
\bibfield{author}{\bibinfo{person}{Yi He}, \bibinfo{person}{Ping Wang},
  \bibinfo{person}{Shiqiang Xiong}, \bibinfo{person}{Chao Chen}, {and}
  \bibinfo{person}{Haixiang Hu}.} \bibinfo{year}{2026}\natexlab{}.
\newblock \showarticletitle{{FinMathBench}: A Formula-Driven Benchmark for
  Evaluating {LLMs}' Math Reasoning Capabilities in Finance}.
\newblock \bibinfo{journal}{\emph{Proceedings of the AAAI Conference on
  Artificial Intelligence}} \bibinfo{volume}{40}, \bibinfo{number}{37}
  (\bibinfo{year}{2026}), \bibinfo{pages}{30987--30995}.
\newblock
\href{https://doi.org/10.1609/aaai.v40i37.40358}{doi:\nolinkurl{10.1609/aaai.v40i37.40358}}


\bibitem[Huang et~al\mbox{.}(2024)]%
        {huang2024openfinllms}
\bibfield{author}{\bibinfo{person}{Jimin Huang}, \bibinfo{person}{Mengxi Xiao},
  \bibinfo{person}{Dong Li}, {et~al\mbox{.}}} \bibinfo{year}{2024}\natexlab{}.
\newblock \bibinfo{title}{{Open-FinLLMs}: Open Multimodal Large Language Models
  for Financial Applications}.
\newblock
\showeprint[arxiv]{2408.11878}~[cs.CL]
\href{https://doi.org/10.48550/arXiv.2408.11878}{doi:\nolinkurl{10.48550/arXiv.2408.11878}}
\newblock
\shownote{Revised June 2025}.


\bibitem[Hull(2023)]%
        {hull2023risk}
\bibfield{author}{\bibinfo{person}{John~C. Hull}.}
  \bibinfo{year}{2023}\natexlab{}.
\newblock \bibinfo{booktitle}{\emph{Risk Management and Financial Institutions}
  (\bibinfo{edition}{6} ed.)}.
\newblock \bibinfo{publisher}{John Wiley \& Sons}, \bibinfo{address}{Hoboken,
  NJ, USA}.
\newblock


\bibitem[Islam et~al\mbox{.}(2023)]%
        {islam2023financebench}
\bibfield{author}{\bibinfo{person}{Pranab Islam}, \bibinfo{person}{Anand
  Kannappan}, \bibinfo{person}{Douwe Kiela}, \bibinfo{person}{Rebecca Qian},
  \bibinfo{person}{Nino Scherrer}, {and} \bibinfo{person}{Bertie Vidgen}.}
  \bibinfo{year}{2023}\natexlab{}.
\newblock \bibinfo{title}{{FinanceBench}: A New Benchmark for Financial
  Question Answering}.
\newblock
\showeprint[arxiv]{2311.11944}
\href{https://doi.org/10.48550/arXiv.2311.11944}{doi:\nolinkurl{10.48550/arXiv.2311.11944}}
\newblock
\shownote{arXiv preprint}.


\bibitem[Larsen et~al\mbox{.}(2026)]%
        {larsen2026vfillm}
\bibfield{author}{\bibinfo{person}{Alicia Larsen}, \bibinfo{person}{Victoire
  Laurent}, \bibinfo{person}{Aulia~Kharis Rakhmasari}, \bibinfo{person}{Lara
  Turgut}, {and} \bibinfo{person}{Nino Antulov-Fantulin}.}
  \bibinfo{year}{2026}\natexlab{}.
\newblock \bibinfo{title}{{V-FiLLM}: Verified Financial {LLM} Reasoning
  Benchmark}.
\newblock
\showeprint[arxiv]{2608.11047}~[cs.AI]
\href{https://doi.org/10.48550/arXiv.2608.11047}{doi:\nolinkurl{10.48550/arXiv.2608.11047}}


\bibitem[Lopez-Lira and Tang(2026)]%
        {lopezlira2026forecast}
\bibfield{author}{\bibinfo{person}{Alejandro Lopez-Lira} {and}
  \bibinfo{person}{Yuehua Tang}.} \bibinfo{year}{2026}\natexlab{}.
\newblock \showarticletitle{Can {ChatGPT} Forecast Stock Price Movements?
  Return Predictability and Large Language Models}.
\newblock \bibinfo{journal}{\emph{Journal of Financial Economics}}
  \bibinfo{volume}{184} (\bibinfo{year}{2026}), \bibinfo{pages}{104335}.
\newblock
\href{https://doi.org/10.1016/j.jfineco.2026.104335}{doi:\nolinkurl{10.1016/j.jfineco.2026.104335}}


\bibitem[Rahimikia and Drinkall(2024)]%
        {rahimikia2024revisiting}
\bibfield{author}{\bibinfo{person}{Eghbal Rahimikia} {and}
  \bibinfo{person}{Felix Drinkall}.} \bibinfo{year}{2024}\natexlab{}.
\newblock \bibinfo{booktitle}{\emph{{Re(Visiting)} Large Language Models in
  Finance}}.
\newblock \bibinfo{type}{{T}echnical {R}eport}. \bibinfo{institution}{SSRN}.
\newblock
\href{https://doi.org/10.2139/ssrn.4963618}{doi:\nolinkurl{10.2139/ssrn.4963618}}
\newblock
\shownote{Working paper; revised August 2026}.


\bibitem[Tang et~al\mbox{.}(2025)]%
        {tang2025financereasoning}
\bibfield{author}{\bibinfo{person}{Zichen Tang}, \bibinfo{person}{Haihong E},
  \bibinfo{person}{Ziyan Ma}, \bibinfo{person}{Haoyang He},
  \bibinfo{person}{Jiacheng Liu}, \bibinfo{person}{Zhongjun Yang},
  \bibinfo{person}{Zihua Rong}, \bibinfo{person}{Rongjin Li},
  \bibinfo{person}{Kun Ji}, \bibinfo{person}{Qing Huang},
  \bibinfo{person}{Xinyang Hu}, \bibinfo{person}{Yang Liu}, {and}
  \bibinfo{person}{Qianhe Zheng}.} \bibinfo{year}{2025}\natexlab{}.
\newblock \showarticletitle{{FinanceReasoning}: Benchmarking Financial
  Numerical Reasoning More Credible, Comprehensive and Challenging}. In
  \bibinfo{booktitle}{\emph{Proceedings of the 63rd Annual Meeting of the
  Association for Computational Linguistics (Volume 1: Long Papers)}},
  \bibfield{editor}{\bibinfo{person}{Wanxiang Che}, \bibinfo{person}{Joyce
  Nabende}, \bibinfo{person}{Ekaterina Shutova}, {and}
  \bibinfo{person}{Mohammad~Taher Pilehvar}} (Eds.).
  \bibinfo{publisher}{Association for Computational Linguistics},
  \bibinfo{address}{Vienna, Austria}, \bibinfo{pages}{15721--15749}.
\newblock
\showISBNx{979-8-89176-251-0}
\showeprint[arxiv]{2506.05828}~[cs.CL]
\href{https://doi.org/10.18653/v1/2025.acl-long.766}{doi:\nolinkurl{10.18653/v1/2025.acl-long.766}}


\bibitem[Tong et~al\mbox{.}(2026)]%
        {tong2026longhorizonfinance}
\bibfield{author}{\bibinfo{person}{Xinke Tong}, \bibinfo{person}{Xuanming
  Zhang}, \bibinfo{person}{Tianyi Tang}, \bibinfo{person}{An Yang},
  \bibinfo{person}{Jiatu Hu}, \bibinfo{person}{Guojie Lin},
  \bibinfo{person}{Zhenzhen Shi}, \bibinfo{person}{Lingfeng Zeng},
  \bibinfo{person}{Boyu Yang}, \bibinfo{person}{Bing Zhao}, \bibinfo{person}{Hu
  Wei}, \bibinfo{person}{Lin Qu}, {and} \bibinfo{person}{Dayiheng Liu}.}
  \bibinfo{year}{2026}\natexlab{}.
\newblock \bibinfo{title}{Are the Financial Reasoning from {LLMs} Credible? {A}
  Real World Test over Long-Horizon Statements}.
\newblock
\showeprint[arxiv]{2607.28661}~[cs.CL]
\urldef\tempurl%
\url{https://arxiv.org/abs/2607.28661}
\showURL{%
\tempurl}


\bibitem[Xie et~al\mbox{.}(2024)]%
        {xie2024finben}
\bibfield{author}{\bibinfo{person}{Qianqian Xie}, \bibinfo{person}{Weiguang
  Han}, \bibinfo{person}{Zhengyu Chen}, \bibinfo{person}{Ruoyu Xiang},
  \bibinfo{person}{Xiao Zhang}, \bibinfo{person}{Yueru He},
  \bibinfo{person}{Mengxi Xiao}, \bibinfo{person}{Dong Li},
  \bibinfo{person}{Yongfu Dai}, \bibinfo{person}{Duanyu Feng},
  \bibinfo{person}{Yijing Xu}, \bibinfo{person}{Haoqiang Kang},
  \bibinfo{person}{Ziyan Kuang}, \bibinfo{person}{Chenhan Yuan},
  \bibinfo{person}{Kailai Yang}, \bibinfo{person}{Zheheng Luo},
  \bibinfo{person}{Tianlin Zhang}, \bibinfo{person}{Zhiwei Liu},
  \bibinfo{person}{Guojun Xiong}, \bibinfo{person}{Zhiyang Deng},
  \bibinfo{person}{Yuechen Jiang}, \bibinfo{person}{Zhiyuan Yao},
  \bibinfo{person}{Haohang Li}, \bibinfo{person}{Yangyang Yu},
  \bibinfo{person}{Gang Hu}, \bibinfo{person}{Jiajia Huang},
  \bibinfo{person}{Xiao-Yang Liu}, \bibinfo{person}{Alejandro Lopez-Lira},
  \bibinfo{person}{Benyou Wang}, \bibinfo{person}{Yanzhao Lai},
  \bibinfo{person}{Hao Wang}, \bibinfo{person}{Min Peng},
  \bibinfo{person}{Sophia Ananiadou}, {and} \bibinfo{person}{Jimin Huang}.}
  \bibinfo{year}{2024}\natexlab{}.
\newblock \showarticletitle{{FinBen}: A Holistic Financial Benchmark for Large
  Language Models}. In \bibinfo{booktitle}{\emph{Advances in Neural Information
  Processing Systems 37 (NeurIPS 2024), Datasets and Benchmarks Track}},
  Vol.~\bibinfo{volume}{37}. \bibinfo{publisher}{Neural Information Processing
  Systems Foundation, Inc. (NeurIPS)}, \bibinfo{address}{Vancouver, BC,
  Canada}, \bibinfo{pages}{95716--95743}.
\newblock
\showeprint[arxiv]{2402.12659}
\href{https://doi.org/10.52202/079017-3033}{doi:\nolinkurl{10.52202/079017-3033}}


\bibitem[Xie et~al\mbox{.}(2023)]%
        {xie2023pixiu}
\bibfield{author}{\bibinfo{person}{Qianqian Xie}, \bibinfo{person}{Weiguang
  Han}, \bibinfo{person}{Xiao Zhang}, \bibinfo{person}{Yanzhao Lai},
  \bibinfo{person}{Min Peng}, \bibinfo{person}{Alejandro Lopez-Lira}, {and}
  \bibinfo{person}{Jimin Huang}.} \bibinfo{year}{2023}\natexlab{}.
\newblock \showarticletitle{{PIXIU}: A Comprehensive Benchmark, Instruction
  Dataset and Large Language Model for Finance}. In
  \bibinfo{booktitle}{\emph{Advances in Neural Information Processing
  Systems}}, Vol.~\bibinfo{volume}{36}. \bibinfo{publisher}{Neural Information
  Processing Systems Foundation, Inc. (NeurIPS)}, \bibinfo{address}{New
  Orleans, LA, USA}, \bibinfo{pages}{33469--33484}.
\newblock
\href{https://doi.org/10.52202/075280-1454}{doi:\nolinkurl{10.52202/075280-1454}}


\bibitem[Xie et~al\mbox{.}(2026)]%
        {xie2026finchain}
\bibfield{author}{\bibinfo{person}{Zhuohan Xie}, \bibinfo{person}{Daniil Orel},
  \bibinfo{person}{Rushil Thareja}, \bibinfo{person}{Dhruv Sahnan},
  \bibinfo{person}{Hachem Madmoun}, \bibinfo{person}{Fan Zhang},
  \bibinfo{person}{Debopriyo Banerjee}, \bibinfo{person}{Georgi~Nenkov
  Georgiev}, \bibinfo{person}{Xueqing Peng}, \bibinfo{person}{Lingfei Qian},
  \bibinfo{person}{Jimin Huang}, \bibinfo{person}{Jinyan Su},
  \bibinfo{person}{Aaryamonvikram Singh}, \bibinfo{person}{Rui Xing},
  \bibinfo{person}{Rania Elbadry}, \bibinfo{person}{Chen Xu},
  \bibinfo{person}{Haonan Li}, \bibinfo{person}{Fajri Koto},
  \bibinfo{person}{Ivan Koychev}, \bibinfo{person}{Tanmoy Chakraborty},
  \bibinfo{person}{Yuxia Wang}, \bibinfo{person}{Salem Lahlou},
  \bibinfo{person}{Veselin Stoyanov}, \bibinfo{person}{Sophia Ananiadou}, {and}
  \bibinfo{person}{Preslav Nakov}.} \bibinfo{year}{2026}\natexlab{}.
\newblock \showarticletitle{{FinChain}: A Symbolic Benchmark for Verifiable
  Chain-of-Thought Financial Reasoning}. In
  \bibinfo{booktitle}{\emph{Proceedings of the 64th Annual Meeting of the
  Association for Computational Linguistics (Volume 1: Long Papers)}}.
  \bibinfo{publisher}{Association for Computational Linguistics},
  \bibinfo{address}{San Diego, California, United States},
  \bibinfo{pages}{14529--14553}.
\newblock
\showeprint[arxiv]{2506.02515}~[cs.CL]
\href{https://doi.org/10.18653/v1/2026.acl-long.662}{doi:\nolinkurl{10.18653/v1/2026.acl-long.662}}


\bibitem[Zhang et~al\mbox{.}(2025)]%
        {zhang2025xfinbench}
\bibfield{author}{\bibinfo{person}{Zhihan Zhang}, \bibinfo{person}{Yixin Cao},
  {and} \bibinfo{person}{Lizi Liao}.} \bibinfo{year}{2025}\natexlab{}.
\newblock \showarticletitle{{XFinBench}: Benchmarking {LLMs} in Complex
  Financial Problem Solving and Reasoning}. In
  \bibinfo{booktitle}{\emph{Findings of the Association for Computational
  Linguistics: ACL 2025}}, \bibfield{editor}{\bibinfo{person}{Wanxiang Che},
  \bibinfo{person}{Joyce Nabende}, \bibinfo{person}{Ekaterina Shutova}, {and}
  \bibinfo{person}{Mohammad~Taher Pilehvar}} (Eds.).
  \bibinfo{publisher}{Association for Computational Linguistics},
  \bibinfo{address}{Vienna, Austria}, \bibinfo{pages}{8715--8758}.
\newblock
\showISBNx{979-8-89176-256-5}
\showeprint[arxiv]{2508.15861}~[cs.CL]
\href{https://doi.org/10.18653/v1/2025.findings-acl.457}{doi:\nolinkurl{10.18653/v1/2025.findings-acl.457}}


\bibitem[Zhu et~al\mbox{.}(2021)]%
        {zhu2021tatqa}
\bibfield{author}{\bibinfo{person}{Fengbin Zhu}, \bibinfo{person}{Wenqiang
  Lei}, \bibinfo{person}{Youcheng Huang}, \bibinfo{person}{Chao Wang},
  \bibinfo{person}{Shuo Zhang}, \bibinfo{person}{Jiancheng Lv},
  \bibinfo{person}{Fuli Feng}, {and} \bibinfo{person}{Tat-Seng Chua}.}
  \bibinfo{year}{2021}\natexlab{}.
\newblock \showarticletitle{{TAT-QA}: A Question Answering Benchmark on a
  Hybrid of Tabular and Textual Content in Finance}. In
  \bibinfo{booktitle}{\emph{Proceedings of the 59th Annual Meeting of the
  Association for Computational Linguistics and the 11th International Joint
  Conference on Natural Language Processing (Volume 1: Long Papers)}}.
  \bibinfo{publisher}{Association for Computational Linguistics},
  \bibinfo{address}{Online}, \bibinfo{pages}{3277--3287}.
\newblock
\showeprint[arxiv]{2105.07624}
\href{https://doi.org/10.18653/v1/2021.acl-long.254}{doi:\nolinkurl{10.18653/v1/2021.acl-long.254}}


\end{thebibliography}



\clearpage
\appendix
\section*{Appendix}

\section{Evaluation Records and Reference Checks}
\label{app:reference-audit}

\noindent\textbf{Evaluation record.}
The evaluation contains 175 Standard and 30 Long-horizon tasks, each scored from one recorded response per model configuration. All eleven configurations have grades for the 30 Long-horizon tasks; Astra and Sol each lack one Standard evaluation (\texttt{risk-119} and \texttt{risk-117}, respectively). As a result, the intersection of evaluated task IDs contains 203 tasks. An additional graph record, \texttt{risk-006}, has no evaluation among the eleven configurations and is not included in accuracy calculations. We preserve the original identifiers for traceability, without treating identifier prefixes as separate task regimes.

Only the grade \texttt{pass} contributes to exact Pass@1. The other recorded statuses (\texttt{partial}, \texttt{near}, \texttt{fail\_close}, and \texttt{fail}) are non-passes rather than ordinal measures of financial-error severity. Response extraction, numerical tolerances, and field-level completeness affect these outcomes. The selected manual reviews help interpret individual cases but do not establish failure-category prevalence. Detailed grade distributions, paired-pass tabulations, and supplementary cases are retained as machine-readable and source artifacts accompanying the project, rather than repeated as manuscript tables.

\noindent\textbf{Reference and scoring checks.}
The audit distinguishes conventional choices from potentially incorrect numerical references. Financial conventions require clarification for the median in \texttt{risk-005}, annualized-return conventions in \texttt{risk-034}, the moving-average signal in \texttt{risk-063}, leverage definitions in \texttt{risk-092}, and the simple-versus-Wilder average-true-range interpretation in \texttt{risk-116}. Numerical checks are still required for the Pearson correlations in \texttt{risk-117}, vector-autoregression information criteria in \texttt{risk-120}, and Jarque--Bera statistics in \texttt{risk-125}. The reported yield matching in \texttt{risk-082} merits a tolerance and extraction audit. Claude's \texttt{risk-126} run reached the 16,384-token output budget, which establishes a completion limitation rather than a specific financial-reasoning failure. These flags do not automatically vindicate divergent model outputs; grades are retained as recorded pending verified corrections.

\section{Illustrative Financial Dependency}
\label{app:illustrative-task}

The Long-horizon task \texttt{long-001} illustrates how individually straightforward calculations can lead to a conflicting pair of financial constraints. Its machine-readable dependency graph contains 31 vertices and 35 directed edges. The following summary traces the key quantities; it is not intended as a full rendering of the graph.

\noindent\textbf{Market shock and collateral.}
A bank holds \$60M of bonds with modified duration 5.2 and faces a 120-basis-point increase in yield. The specified duration approximation gives a \$3.744M loss, reducing collateral value to \$56.256M. The \$48M secured borrowing requires 125\% collateral coverage, creating a \$3.744M margin call:
\begin{align*}
\text{Bond loss} &= 60(5.2)(0.012) = 3.744,\\
\text{Margin call} &= 1.25(48) - 56.256 = 3.744.
\end{align*}

\noindent\textbf{Funding response and constraint check.}
Paying the call reduces the bank's \$4M cash balance to \$0.256M. It draws \$1.244M on an emergency credit line to restore the \$1.5M cash floor. That financing decision raises total debt to \$80.744M, resulting in
\begin{align*}
\text{Required draw} &= 1.5 - 0.256 = 1.244,\\
\frac{\text{Debt}}{\text{Equity}} &= \frac{80.744}{20} = 4.0372.
\end{align*}

The cash-floor condition is satisfied, but the debt-to-equity ratio exceeds the permitted \(4.0\times\) ceiling. The example shows why a valid financial-risk conclusion must propagate the effects of a liquidity remedy to subsequent leverage constraints: restoring compliance at one stage can create noncompliance at another.

\section{Evaluated Model Specifications and Profiles}
\label{sec:appendix-models}

This appendix provides architectural specifications, release metadata, and primary target workloads for all evaluated models.

\begin{table*}[t]
\centering
\caption{Summary of Evaluated Language Models}
\label{tab:evaluated-models}
\begin{tabular}{llrll}
\toprule
\textbf{Model} & \textbf{Developer / Family} & \textbf{Parameters} & \textbf{Release Date} & \textbf{Primary Specialization} \\
\midrule
Claude Sonnet 5.5 & Anthropic & Undisclosed & Sep 28, 2026 & Agentic coding, vision, multi-step workflows \\
GPT-6 Astra & OpenAI & Undisclosed & Sep 2026 & Computer use, long-horizon agents, deep research \\
GPT-6 Sol & OpenAI & Undisclosed & Sep 2026 & Complex reasoning, long-context tool use \\
GPT-6 Luna & OpenAI & Undisclosed & Sep 2026 & High-throughput extraction, summarization \\
Gemini 3.8 Flash & Google & Undisclosed & Sep 2026 & Autonomous agents, fast enterprise pipelines \\
Gemini 3.1 Pro Preview & Google & Undisclosed & Feb 2026 & Frontier reasoning, multimodal, finance/sheets \\
Phi-4-Reasoning-Plus & Microsoft Research & 14B & Apr 2025 & CoT reasoning in memory-constrained setups \\
DeepSeek-R1-Distill-32B & DeepSeek & 32B & Early 2025 & Distilled mathematical and logical reasoning \\
Fin-o1-8B & TheFinAI & 8B & Feb 2025 & Financial reasoning and domain logic \\
DianJin-R1-7B & DianJin & 7B & Early 2025 & Structured reasoning (SFT + GRPO) \\
Qwen2.5-7B-Instruct & Alibaba & 7B & Mid 2025 & General instruction, multilingual, code \\
\bottomrule
\end{tabular}
\end{table*}

\subsection{Detailed Model Profiles}

\subsubsection{Claude Sonnet 5.5}
Claude Sonnet 5.5 is Anthropic's second release in the Claude 5.5 generation, announced on September 28, 2026 (\url{https://www.anthropic.com/claude-sonnet-5-5}). Its exact parameter count has not been officially disclosed. The model is engineered to operate over 30\% faster and cost up to 30\% less than Claude Sonnet 5 across common enterprise tasks, exhibiting specialized capabilities in agentic coding, extended multi-step workflows, long-horizon software engineering, and multimodal visual analysis.

\subsubsection{GPT-6 Astra}
GPT-6 Astra is OpenAI's flagship frontier model released in September 2026 (\url{https://developers.openai.com/api/docs/models/gpt-6-astra}), with its exact parameter count undisclosed. Positioned above Sol and Luna in the GPT-6 family, it is engineered for the most demanding end-to-end workloads. Astra exhibits specialized, state-of-the-art capabilities in computer and browser use, long-horizon agentic operations, multi-step software engineering, and deep research across extensive context windows.

\subsubsection{GPT-6 Sol}
GPT-6 Sol is a flagship OpenAI foundation model released in September 2026 (\url{https://azure.microsoft.com/en-us/blog/gpt-6-astra-sol-and-luna-for-production-agents-in-microsoft-foundry/}) with undisclosed parameter counts. Positioned as the architectural successor to GPT-5.6 Sol, it is optimized for high-complexity knowledge synthesis and autonomous production agents, demonstrating heightened fidelity in code refactoring, prolonged debugging tasks, and tool-augmented multi-step reasoning over wide context windows.

\subsubsection{GPT-6 Luna}
GPT-6 Luna represents the low-latency companion architecture to GPT-6 Sol within the OpenAI lineup, released in September 2026 (\url{https://docs.aws.amazon.com/bedrock/latest/userguide/model-card-openai-gpt-6-luna.html}) with an undisclosed parameter size. Tailored specifically for cost-effective, high-throughput execution at enterprise scale, it targets structured entity extraction, document summarization, operational routing, and standard conversational interactions where compute-intensive multi-hop reasoning is unnecessary.

\subsubsection{Gemini 3.8 Flash}
Gemini 3.8 Flash is Google's primary high-efficiency reasoning engine, launched in September 2026 (\url{https://ai.google.dev/gemini-api/docs/models/gemini-3.8-flash}) with parameter weights not publicly disclosed. It balances low operational latency with deep analytical capabilities, targeting fast autonomous agent execution, high-frequency enterprise backend pipelines, and continuous software engineering tasks.

\subsubsection{Gemini 3.1 Pro Preview}
Gemini 3.1 Pro Preview is Google's multimodal frontier reasoning model made available in February 2026 (\url{https://openrouter.ai/google/gemini-3.1-pro-preview}), featuring an undisclosed parameter scale. The system natively integrates visual processing, native function execution, schema-constrained outputs, and real-time retrieval grounding, targeting complex quantitative domains including spreadsheet automation, econometric modeling, and multimodal analysis.

\subsubsection{Phi-4-Reasoning-Plus}
Phi-4-Reasoning-Plus is an open-weight, 14-billion-parameter dense decoder-only model released by Microsoft Research in April 2025 (\url{https://huggingface.co/microsoft/Phi-4-reasoning-plus}). Fine-tuned from the Phi-4 base architecture via chain-of-thought supervised trajectories and reinforcement learning, it delivers enhanced mathematical and logical reasoning within resource-constrained execution environments.

\subsubsection{DeepSeek-R1-Distill-32B}
DeepSeek-R1-Distill-32B is a 32-billion-parameter open-weight reasoning model released in early 2025 (\url{https://huggingface.co/deepseek-ai/DeepSeek-R1-Distill-Qwen-32B}). Built by distilling reasoning outputs from the full DeepSeek R1 model into a Qwen foundational backbone, it retains high mathematical, scientific, and deductive reasoning performance while maintaining a footprint deployable on consumer-tier accelerator hardware.

\subsubsection{Fin-o1-8B}
Fin-o1-8B is an 8-billion-parameter research model developed by TheFinAI and released in February 2025 (\url{https://huggingface.co/TheFinAI/Fin-o1-8B}). Derived as a specialized fine-tune of Qwen3-8B, the architecture is tailored explicitly toward corporate finance reasoning, fiscal statement parsing, and specialized quantitative arithmetic tasks.

\subsubsection{DianJin-R1-7B}
DianJin-R1-7B is an open-source, 7-billion-parameter language model published in early 2025 (\url{https://huggingface.co/DianJin/DianJin-R1-7B}). Derived from Qwen2.5-7B-Instruct, it was trained via a two-stage alignment pipeline comprising Supervised Fine-Tuning (SFT) over reasoning trajectories followed by Group Relative Policy Optimization (GRPO) reinforcement learning to reinforce structural consistency and factual precision.

\subsubsection{Qwen2.5-7B-Instruct}
Qwen2.5-7B-Instruct is an instruction-tuned, 7-billion-parameter foundational model created by Alibaba and released in mid 2025 (\url{https://huggingface.co/Qwen/Qwen2.5-7B-Instruct}). Pre-trained on an 18-trillion token multilingual corpus and incorporating architectural mechanisms such as Grouped Query Attention (GQA) and SwiGLU activations, it delivers competitive general-purpose baseline performance across multilingual translation, coding, and mathematical reasoning.

\section{Illustrative Evaluation Example}
\label{app:evaluation-example}

To illustrate the end-to-end evaluation, we present one successful
Long-horizon task from the recorded evaluation. This example is selected
for its interconnected market, liquidity, funding, and capital constraints;
it is illustrative rather than representative of average performance.
The stored task record contains a natural-language scenario, a reference
solution, and a prescribed output template. The response was generated by \texttt{gpt-6-astra} in one recorded attempt
(\emph{pass@1}).

\paragraph{Task.}
Task \texttt{long-009} presents the following scenario. The text below
reproduces the substantive task prompt; the common instruction to fill
an exact JSON template is not repeated because the populated response
is given in full immediately afterward.

\begin{quote}
\small
A bank has \$100M of total assets, \$10M of equity, and \$3M of cash. It pledges a \$50M bond portfolio against \$38M of repo borrowing. The bonds have modified duration 4.0. Market yields rise by 100 basis points; ignore convexity. At the same time, the repo haircut rises from 20\% to 25\%.

After the shock, required collateral market value equals repo borrowing divided by (1\ensuremath{-}haircut). Any collateral shortfall is met with a cash margin payment. The bank must maintain at least \$1M of cash; if the margin payment would push cash below that level, it borrows unsecured funds just sufficient to restore the \$1M minimum.

The bond loss reduces both assets and equity. New unsecured borrowing adds cash and liabilities but does not change equity. The maximum assets/equity leverage ratio is 12\ensuremath{\times}. Assume any equity injection is retained as cash and therefore increases both assets and equity dollar-for-dollar.

Calculate the bond loss and stressed collateral value, new collateral requirement, margin call, cash after the call, unsecured funding needed, post-funding assets/equity and leverage, and the minimum equity injection required to restore leverage to 12\ensuremath{\times}.
\end{quote}

\paragraph{Recorded model response.}
The model first connected the interest-rate shock to the duration-based
bond loss, then recalculated the repo collateral requirement, determined
the cash shortfall and unsecured borrowing, and finally solved for the
cash-retained equity injection needed to satisfy the leverage limit.
Its structured answer, reproduced without numerical rounding or edited
fields, was:

\begingroup\footnotesize
\begin{verbatim}
{
 "bond_loss_m": 2,
 "stressed_collateral_value_m": 48,
 "new_collateral_requirement_m": 50.6666666667,
 "margin_call_m": 2.6666666667,
 "cash_after_call_m": 0.3333333333,
 "unsecured_funding_needed_m": 0.6666666667,
 "post_funding_equity_m": 8,
 "post_funding_assets_m": 98.6666666667,
 "post_funding_leverage_ratio": 12.3333333333,
 "minimum_equity_injection_m": 0.2424242424,
 "post_equity_cure_leverage_ratio": 12
}
\end{verbatim}
\endgroup

The accompanying final explanation in the recorded response read:
\begin{quote}
\small
All dollar amounts are in millions. The 100-basis-point yield increase causes an approximate bond loss of 50 \ensuremath{\times} 4 \ensuremath{\times} 0.01 = 2, leaving stressed collateral worth 48. At the new 25\% haircut, required collateral is 38 / 0.75 = 50.666667, so the cash margin call is 2.666667. Available cash after the call is 0.333333, requiring unsecured borrowing of 0.666667 to restore the minimum cash balance of 1. Posted cash margin remains an asset rather than an expense. Consequently, post-funding assets are 100 \ensuremath{-} 2 + 0.666667 = 98.666667, and equity is 10 \ensuremath{-} 2 = 8. Leverage is therefore 12.333333 times. Since an equity injection increases both assets and equity, the minimum injection is (98.666667 \ensuremath{-} 12 \ensuremath{\times} 8) / 11 = 0.242424, approximately \$242,424.24. The resulting leverage ratio is exactly 12 times.
\end{quote}

\paragraph{Reference verification and grading.}
The benchmark reference independently specifies all 11 quantities.
The main financial dependencies can be checked directly (dollar amounts
in millions):
\begin{align*}
\text{Bond loss} &= 50(4)(0.01)=2,\\
\text{Collateral required} &= 38/(1-0.25)=50.6667,\\
\text{Cash margin call} &= 50.6667-48=2.6667,\\
\text{Unsecured funding} &= 1-(3-2.6667)=0.6667,\\
\text{Post-funding leverage} &= 98.6667/8=12.3333,\\
\text{Equity injection} &= \frac{98.6667-12(8)}{12-1}
\approx 0.242424.
\end{align*}

As specified in the task, posted cash collateral remains an asset;
the margin transfer therefore does not itself reduce total assets.
The bond-price loss reduces assets and equity by \$2 million, whereas
new unsecured financing increases assets and liabilities without
increasing equity. An equity injection retained as cash raises both
assets and equity, producing the requested 12-times leverage ratio.

For this single response, the evaluator reported
\texttt{status=pass}, \texttt{matched=11/11}, and
\texttt{schema\_ok=true}. Every required reference field was matched
under the benchmark's strict numerical comparison, and the recorded
response respected the prescribed output schema. This is a complete
task-level pass, not partial credit from isolated correct numbers.

\section{Financial Domain Coverage and Annotated Difficulty}
\label{app:domain-coverage}

Table~\ref{tab:domain-coverage} summarizes the primary financial domains
represented in the evaluated benchmark: 175 Standard tasks and 30
Long-horizon tasks. Domains refer to the primary task annotation rather
than mutually exclusive financial reasoning skills; a task can involve
multiple concepts while contributing to only one domain count. We report
an authored difficulty score alongside the counts to describe the
composition of the task set, not to estimate model performance.

\begin{table*}[t]
\centering
\caption{Financial-domain distribution and authored difficulty for all 205
RiskReason tasks. Difficulty is the sum of eight task-annotation scores;
``Median [range]'' reports the median and minimum--maximum within each
domain. Shares are individually rounded to one decimal, and reasoning
drivers are illustrative rather than exhaustive.}
\label{tab:domain-coverage}
\begingroup
\footnotesize
\setlength{\tabcolsep}{2.7pt}
\renewcommand{\arraystretch}{1.14}
\begin{tabular*}{\textwidth}{@{\extracolsep{\fill}}p{0.232\textwidth}rrrrrp{0.11\textwidth}p{0.255\textwidth}@{}}
\toprule
\textbf{Financial domain} & \textbf{Std.} & \textbf{Long} & \textbf{Total} & \textbf{\%} & \textbf{Mean} & \textbf{Median [range]} & \textbf{Illustrative reasoning drivers} \\
\midrule
Credit Risk & 29 & 7 & 36 & 17.6 & 16.00 & 16 [6--24] & Default probabilities; LGD; expected loss \\
Investments \& Portfolio Management & 31 & 1 & 32 & 15.6 & 8.66 & 8 [4--20] & Return statistics; portfolio measures \\
Banking \& ALM & 22 & 6 & 28 & 13.7 & 15.61 & 17.5 [6--23] & Interest-rate gaps; liquidity; leverage \\
Market Risk & 19 & 3 & 22 & 10.7 & 12.77 & 13 [4--19] & VaR; volatility; EWMA series \\
Corporate Finance & 15 & 6 & 21 & 10.2 & 13.33 & 12 [4--20] & WACC; capital budgeting; cash flows \\
Counterparty Credit Risk & 9 & 3 & 12 & 5.9 & 17.08 & 18 [9--23] & Exposure modeling; collateral conditions \\
Derivatives & 12 & 0 & 12 & 5.9 & 11.50 & 11.5 [7--17] & Option valuation; payoff scenarios \\
Capital Markets \& Trading & 8 & 1 & 9 & 4.4 & 11.33 & 10 [5--20] & Indicators; signal and execution rules \\
Fixed Income & 9 & 0 & 9 & 4.4 & 8.56 & 8 [3--13] & Bond pricing; yields; discounting \\
Enterprise Risk Management & 8 & 0 & 8 & 3.9 & 20.25 & 20 [17--27] & Cross-risk dependencies; conditional logic \\
Insurance \& Actuarial & 6 & 1 & 7 & 3.4 & 13.00 & 11 [9--20] & Actuarial values; cash-flow chains \\
Retirement \& Pensions & 3 & 1 & 4 & 2.0 & 9.50 & 7 [5--19] & Annuities; contribution decisions \\
Operational Risk & 2 & 0 & 2 & 1.0 & 15.50 & 15.5 [9--22] & Scenario analysis; branching rules \\
Regulatory \& Capital Risk & 1 & 1 & 2 & 1.0 & 17.50 & 17.5 [17--18] & Capital ratios; multi-step regulatory rules \\
Model Risk & 1 & 0 & 1 & 0.5 & 20.00 & 20 [20--20] & Structured validation; conditional logic \\
\midrule
\textbf{Total} & \textbf{175} & \textbf{30} & \textbf{205} & \textbf{100.0} & --- & --- & --- \\
\bottomrule
\end{tabular*}
\endgroup
\end{table*}

The difficulty scores are metadata derived from annotated reasoning
requirements (depth, structure, representation, temporal, probabilistic,
conditional, unit-conversion, and domain-knowledge components). They
should not be read as empirically calibrated difficulty or as independent
measurements of financial expertise. Domain-level averages are
particularly sensitive to individual tasks in small categories.

\end{document}